\documentclass[format=acmsmall, review=false, screen=true]{acmart}

\usepackage{booktabs} % For formal tables
\usepackage[noend]{algpseudocode}
\usepackage{tabularx}
\usepackage{tikz,pgfplots}
\usepackage[ruled]{algorithm2e} % For algorithms

\SetAlFnt{\small}
\SetAlCapFnt{\small}
\SetAlCapNameFnt{\small}
\SetAlCapHSkip{0pt}
\IncMargin{-\parindent}

\usepackage{etoolbox}
\makeatletter
\patchcmd{\maketitle}{\@copyrightspace}{}{}{}
\makeatother

\renewcommand\footnotetextcopyrightpermission[1]{} % removes footnote with conference information in first column
\acmDOI{}

\date{}

\acmVolume{0}
\acmNumber{0}
\acmArticle{0}
\acmYear{0}
\acmMonth{0}
\begin{document}
% Title portion. Note the short title for running heads
\title[The Multi-Event-Class Synchronization (MECS) Algorithm]{The Multi-Event-Class Synchronization (MECS) Algorithm}

\author{Paolo Alborno}
\affiliation{%
\institution{University of Genova}
}
\author{Gualtiero Volpe}
\affiliation{%
	\institution{University of Genova}
}
\author{Maurizio Mancini}
\affiliation{%
	\institution{University of Genova}
}
\author{Radoslaw Niewiadomski}
\affiliation{%
	\institution{University of Genova}
}
\author{Stefano Piana}
\affiliation{%
	\institution{University of Genova}
}
\author{Antonio Camurri}
\affiliation{%
	\institution{University of Genova}
}

\begin{abstract}
Synchronization is a fundamental component of computational models of human behavior, at both intra-personal and inter-personal level. Event synchronization analysis was originally conceived with the aim of providing a simple and robust method to measure synchronization between two time series. In this paper we propose a novel method extending the state-of-the-art of the event synchronization techniques: the Multi-Event-Class Synchronization (MECS) algorithm. MECS measures the synchronization between relevant events belonging to different event classes that are detected in multiple time series. Its motivation emerged from the need to model non-verbal multimodal signals in Human-Computer Interaction.
Using MECS, synchronization can be computed between events belonging to the same class (intra-class synchronization) or between events belonging to different classes (inter-class synchronization). In the paper we also show how our technique can deal with macro-events (i.e., sets of events satisfying constraints) and macro-classes (i.e., sets of classes).
In the last part of the paper, we apply the proposed method to two types of data i) artificial and 2) real-world case study concerning analysis of human multimodal behavior.
\end{abstract}




\maketitle

\renewcommand{\shortauthors}{P.Alborno et al.}

\section{Introduction}
This paper presents \emph{Multi-Event-Class Synchronization} (\emph{MECS}), a new algorithm to measure the amount of synchronization between events detected in two or more time series. MECS belongs to a family of Event Synchronization (ES) techniques and it is inspired by the work of R. Q. Quiroga and colleagues \cite{quiroga2002event}. The techniques to perform ES analysis are also known under the name of \textit{Measures of spike train synchrony} \cite{kreuz2011measures}. ES analysis is performed for measuring the degree of synchronization between events occurring in a set of time series. The term ``events'' denotes a significant behavior for a system. 

With respect to other existing techniques, MECS brings substantial extensions which allow to model a large set of real-life phenomena. First of all, it deals with multiple classes of events. After grouping events into classes, synchronization is computed within a class, i.e., between events belonging to the same class (\emph{intra-class synchronization}) and between classes, i.e., between events belonging to different classes (\emph{inter-class synchronization}). 
Additionally, events can be combined in \emph{macro-events} on which synchronization is measured. Each \emph{macro-event} is an aggregation of the events that satisfy some constraints. A relevant example of macro-event is a sequence of events. 
Events and macro-events can be grouped in \emph{macro-classes} and synchronization can be computed within and between them.

While many of the existing ES algorithms (e.g., \cite{quiroga2002event,kreuz2009measuring} were developed in the context of brain signal analysis, MECS was created with the purpose of studying multimodal human-human and human-machine interaction. MECS can be applied to a large variety of problems and, in particular, it can be used by human centered systems, multi-modal interfaces for human-machine interaction, or to study multi-modal expressive behaviors of individuals as well as social signals in groups.
%with the long-term goal of endowing machines with the capability to ``decode'' human behaviors.\\
%Studies on the synchronization of expressive human behaviors have a long tradition: 
Indeed, intra-personal synchronization between expressive behaviors of one or more body modalities is an important cue of several emotion displays \cite{Keltner1995,niewiadomski2011}, synchronization between physiological signals and movement kinematics allows one to distinguish between different qualities of human full-body movement \cite{lussu2016}. At the same time, interpersonal synchronization of expressive behaviors in a group of people is an important cue of group cohesion \cite{lakens2011,hung2010} and soft-entrainment \cite{alborno2016}. 

This paper is organized as follows: in Section \ref{sec:beyond} we describe background and present related work. Section \ref{sec:mecs} describes the MECS algorithm. In Section \ref{sec:manipulations} we discuss  \emph{macro-events} and \emph{macro-classes}. Section \ref{sec:impl} presents a simple implementation of the MECS algorithm with pseudo-code, then Section \ref{sec:examples} provides examples of the algorithm execution over a few synthetic sample signals. Finally in Section \ref{sec:case}, we discuss a concrete case study in the framework of human behavior understanding: intra-personal synchronization between two sensory modalities for analysis of expressive behavior. 
%In particular, we analyze multi-modal data consisting of respiration patterns and 3D motion capture data of a person performing two different classes of full-body movements. 

\begin{table}[h]
	\centering
	\caption{Comparison between existing event synchronization algorithms \cite{quiroga2002event, iqbal, kreuz2009measuring} and MECS.}
	
	\begin{tabular}{|r|l|l|l|l|}
		\hline
		\textbf{Reference} & \multicolumn{2}{ c |}{Number of} & \multicolumn{2}{ c |}{Macro} \\
		& \textbf{time series} & \textbf{classes} & \textbf{events} & \textbf{classes} \\
		\hline
		\cite{quiroga2002event} & 2 & 1  & not handled & not handled \\
		\hline
		\cite{iqbal} & M  & N  & not handled & not handled \\
		\hline
		\cite{kreuz2009measuring} & M & 1  & not handled & not handled \\
		\hline
		MECS & M & N & handled & handled \\
		\hline	
	\end{tabular}
	\label{tab:comparison}
\end{table}

\section{Related work}\label{sec:beyond}

Event Synchronization techniques measure synchronization and time-delay patterns between two time series. An example of ES technique was proposed by Quiroga et al. \cite{quiroga2002event}. Given two time series of $n$ samples $x = {x_1,...,x_n}$ and $y = {y_1,...,y_n}$, $x_i, y_i \in R$, 
events are detected and the time instants $t^x_i$ and $t^y_i$, at which events occur in time series $x$ and $y$ respectively, are computed.   
%ES is provided as input with two binary time series $x$ and $y$ (i.e., containing
%sequences of zeros and ones) and measures their synchronization.
%Each item of $x$ and $y$ 
%A binary event is defined as follows:
%
%\begin{equation}
%e =
%\begin{cases}
%0, & \text{event~not~occurred} \\
%1, & \text{event~occurred}
%\end{cases}
%\end{equation}
%
%\noindent
%Then, from the two time series we obtain two vectors $T^x$ and $T^y$ 
%containing the time instants binary events occur at:
%\begin{equation}
%V = \{e_i | i=0,...,n-1\}
%\end{equation}
%
%\noindent
%where $e_i$ is a binary event.
%Given two event vectors $V_x$ and $V_y$ with respectively $X$ and $Y$ elements, 
Synchronization $Q_\tau$ between the detected events is measured as:
\begin{equation}\label{ES_SI}
Q_\tau = \frac{c^\tau(y|x)+c^\tau(x|y)}{\sqrt{m_x m_y}} 
\end{equation}

\noindent
where $c^\tau(y|x)$ and $c^\tau(x|y)$
are, the number of times an event in time series $y$
appears within a time interval defined by parameter $\tau$ after an event appears in time series $x$, and
vice-versa. $m_x$ is the number of events detected in time series $x$ and $m_y$ is the number of events detected in time series $y$. $c(x|y)$ is computed as:

\begin{equation}\label{ES_c}
{c}^{\tau}(x|y) = \sum_{j=1}^{{m}_{x}}\sum_{i=1}^{{m}_{y}}{J}^{\tau}_{ij} 
\end{equation}

\noindent
where ${J}^{\tau}_{ij}$ is defined as follows:  

\begin{equation}\label{ES_J}
{J}^{\tau}_{ij}  = 
\begin{cases} 
1 & \mbox{if } 0 < {t_i}^{x} - {t_j}^{y} \leq{\tau} \\ 
1/2 & \mbox{if } {t_i}^{x} = {t_j}^{y} \\
0 & \mbox{otherwise}
\end{cases}
\end{equation}

\noindent
%While looking counter-intuitive, 
In particular, the second line of Equation \ref{ES_J} is needed, because when two events occur simultaneously they will be counted 2 times by Quiroga's algorithm, a total contribution of 1 to the computation of $Q_\tau$. Also, we highlight that the quantity $Q_\tau$ can be greater than 1. 
%The MECS algorithm, besides extending the ES algorithm of Quiroga et al. with other concepts such as macro classes and events, is designed to provide as output a synchronization value in the $[0,1]$ interval (see Section \ref{sec:mecs}.

%\noindent
%$J$ describes if an event is considered to appear \textit{shortly}
%after another event evaluating if the temporal distance is
%less than a constant threshold $tau$.

Iqpal \& Riek \cite{iqbal} proposed an extension of Quiroga's algorithm to deal with multiple types of events.
In their approach, given two time series $x$ and $y$ and the event type $E_i, i=1...K$ they first compute the synchronization index $Q_\tau(E_i)$ for all events of type $E_i$ using the Equations \ref{ES_SI} - \ref{ES_J}. Next, they compute synchronization of multiple types of events between $x$ and $y$ as the average of $Q_\tau(E_i)$, weighted by the number of events of type $E_i$.
In the last step, they consider $M$ different time series $z_1, ... z_M$ and they compute their pairwise and overall synchronization.
In particular, the individual synchronization of a given time series $z_i$ is the average of all the pairwise synchronizations between $z_i$ and each time series $z_j : i \neq j$ that are beyond a certain fixed threshold $Q_{tresh}$. The overall synchronization is the average of the products of the individual synchronization indexes multiplied by their connectivity values, where the connectivity value is the number of time series pairs having a pairwise synchronization above the threshold $Q_{tresh}$, divided by the total number of pairs.

Kreuz et al. \cite{kreuz2009measuring} presented a multivariate extension to measure the synchrony 
% case to measure the degree of synchrony 
from the relative number of spike occurrences. 

The Multi-Event Class Synchronization (MECS) algorithm we present in this paper introduces new characteristics that are missing in works
%we present in this paper is an extension of the works 
byQuiroga et al., Iqpal \& Riek, and Kreuz et al., as highlighted in Table \ref{tab:comparison}. The table provides a comparison between these algorithms in terms of: i) the maximum number of input time series; ii) the maximum number of classes; iii) the possibility of handling macro-events, 
%such as for example those induced by temporal constraints (sequences of events), 
and iv) macro-classes. 

The MECS algorithm can compute the synchronization between $M=2$ time series and a single event class ($N=1$). Differently from Quiroga et al., it provides a normalized output in $[0,1]$ (see the above comment on the outputs of Quiroga's algorithm).
%is designed to provide as output a synchronization value 
It can also compute synchronization of $M>2$ multivariate time series, but, unlike Kreuz et al. which considers only single event class ($N=1$), it manages multiple classes of events ($N>1$) within the time series.
%If applied on time series pairs (of a set of $M>2$ time series) and $N>1$ event classes, it provides the class by class synchronization (i.e., intra-class synchronization between events of the same class). 
Additionally to Iqpal \& Riek, it manages the computation of synchronization between events belonging to different classes (i.e., inter-class synchronization).

%Unlike the above works, the main contribution of the MECS algorithm is the computation of synchronization between events belonging to different classes (i.e., inter-class synchronization). As \cite{kreuz2009measuring}, MECS also deals with multivariate time series, but it manages multiple classes of events within multivariate time series. 
Finally, differently from all the algorithms mentioned above, MECS introduces the computation of synchronization between $N>1$ event classes over $M\geq2$ time series, handling %manipulations of the events as 
\textit{macro-events} and 
%of the event class set as 
\textit{macro classes}.

\section{Multi-Event Class Synchronization}
\label{sec:mecs}
Multi-Event Class Synchronization (MECS) computes the amount of synchronization between events occurring in a set of time series. Events may belong to different classes: MECS computes (i) a separate synchronization index for each class (\emph{intra-class synchronization}), (ii) a synchronization index for a specific aggregation of classes (\emph{inter-class synchronization}), and a global synchronization index for all classes. 
%\\We will present the how this algorithm can be used to assess different typologies of sub-problems through the an example of study scenario to which  we will refer throughout Sections \ref{sec:mecs} and \ref{sec:manipulations}.

To present MECS, we introduce an example scenario, which is helpful to explain the major features of the algorithm. For example, suppose that we are interested in measuring the level of motor coordination between the members of $N$ users' group performing a motor task (e.g., a fitness exercise). 
A measure of coordination is obtained by evaluating the amount of synchronization between the movements of the users.
Let us consider:

\begin{itemize}
	\item A set $T$ of $N$ time series: $T \equiv{\{TS_1,...TS_N\}}$ 
	\item A set $E$ of $K$ event classes: $E \equiv{\{E_1,...,E_K\}}$
\end{itemize}

%A time series is a sequence of discrete-time data that
% of the real world. 
\noindent  
Time series contain information about significant events in the data. In other words, it describes occurrences of certain phenomena. Introducing the events means to fit a continuous information (e.g., a velocity of a joint) to the discrete domain.
%(often events are equivalent to time stamps).
An event can be, for instance, a local maximum/minimum of the continuous signal (e.g., joints velocity). Such events can be identified automatically (see, e.g., \cite{alborno2016,lussu2016}) or annotated manually.
Events can belong to several classes. %For example, they can describe different behavior changes in the data, such as peaks and valleys. 
Each class $E_1,...,E_K$ characterizes a different type of event, e.g.,  
class $E_1$ might correspond to local maxima of $s$, while $E_2$  to local minima of the same $s$. 
%i.e., it represents something relevant. 
%that can either be detected or annotated within the time series.

Coming back to our example, the time series $TS_1,...TS_N$ can be used describe the motor activity of $N$ users (i.e., time series $TS_j$ models the motor activity of user $j$) and contain events of any class $E_1,...,E_K$ while each class  identifies a specific movement (for example ``step performed'', ``object grabbed'', ``object released'', and so on).

%In the next section we present two different applications and methods of operation of the MECS algorithm to measure the synchronization between the same class of events (\textit{intra-class}) and between different class of events (\textit{inter-class}) on a set of time series.

\subsection{Intra-class synchronization}
\label{subsec:intra}
MECS relies its computation on the temporal distances between events. 
As in \cite{quiroga2002event,kreuz2011measures}, the computation consists of two steps: 
i) the algorithm first detects events coincidences of two different time series in a specific time interval (\emph{coincidence detection}) and counts them, ii) then the number of detected coincidences is normalized with respect to the total number of possible coincidences (\emph{normalization}). 
\noindent
We associate to each event $h$ of class $E_k \in E$, occurring in time series $TS_n \in T$, its occurrence time:
\begin{equation}
\begin{aligned}
t^{n,k}_h \qquad h=1,..., m_{n,k}
\end{aligned}
\end{equation}

\noindent
where $m_{n,k}$ represents the total number of detected events of class $E_k$ occurring in the $n$-th time series $TS_n$. For example, $t^{2,3}_4$ represents the time at which the fourth event belonging to the third class, i.e., $E_3$, occurred in $TS_2$. In the coincidences detection phase, for each pair of time series $\langle TS_{i}\ ;\ TS_{j}\rangle$, the MECS algorithm computes the \textit{amount of coincidence} ${c}_{k}(x,y)$, between an event $x$ detected on time series $TS_i$ and another event $y$ (of the same class of $x$) detected on time series $TS_{j}$ (with $i \neq j$) by measuring the extent to which they are close in time within a certain interval $\tau_k$ (\emph{coincidence window}) that depends on the class of the events.
The temporal distance $d$ between events $x$ and $y$ is computed as:

\begin{equation}
\label{dist}
d(x,y) = |{t^{i,k}_{x}} - {t^{j, k}_{y}}|
\end{equation}
\noindent
The amount of coincidence $c_{k}$ between $x$ and $y$ is defined 
as follows:

\begin{equation}
\label{amountOfCoinc}
{c}_{k}(x,y)  = 
\begin{cases} 
1- \frac{d(x,y)}{\tau_k} & \mbox{if \quad} 0 \leq d(x,y) \leq \tau_k \\ 
0 & \mbox{otherwise}
\end{cases}
\end{equation}

\noindent
Different than \cite{quiroga2002event,kreuz2011measures}, where coincidence is only detected, in our algorithm it is also quantified, such that ${c}_{k}(x,y) \in [0, 1]$. The dimension of the coincidence window $\tau_k$ is either empirically chosen for a specific problem
%or estimated from the dynamics of the phenomenon the specific class of events refer to, 
or it is automatically calculated, e.g., for each pair of events $x$ and $y$ as proposed in \cite{kreuz2011measures}, i.e.: 

\begin{equation}
\tau^{x, y}_k = \frac{1}{2} \dot \min \{ t^{i, k}_{x + 1} - t^{i, k}_{x}, t^{i, k}_{x} - t^{i, k}_{x - 1},
t^{j, k}_{y + 1} - t^{j, k}_{y}, t^{j, k}_{y} - t^{j, k}_{y - 1} \} \nonumber
\label{tau}
\end{equation}

For each class $E_k$, the \textit{overall coincidence} $C_k(i|j)$ of all the events of class $E_k$ in time series $TS_{i}$ with respect to the events of the same class $E_k$ in time series $TS_{j}$ is computed as follows:
\begin{itemize}
	\item First, the average of the ${c}_{k}$ of each event $x$ in time series $TS_{i}$ with respect to all events in time series $TS_{j}$ is calculated;
	\item Then the sum of the average of all the events in time series $TS_{i}$ is taken.
\end{itemize}

\noindent
That is:

\begin{equation}
{C}_{k}(i|j) = 	
\sum\limits_{x=1}^{m_{i,k}}{
	\frac{1}{m_{j,k}} 
	\bigg[ \sum\limits_{y=1}^{m_{j,k}}{
		{c}_{k}(x,y)
	} \bigg]
}
\label{S1}
\end{equation}

\noindent
Equation \ref{S1} shows that event $x$ in $TS_i$ can contribute to the overall coincidence by being coincident (at different extents) with more than one event in $TS_j$. 
Multiple coincidences are usually unwanted and the computation of the coincidence window, as in Equation \ref{tau}, is often performed with the exact purpose of minimizing the likelihood of counting multiple coincidences in one coincidence window $\tau_k$. Since MECS enables to weight coincidences so that a perfect coincidence has a weight of 1.0, and the amount of coincidence decreases along the coincidence window $\tau_k$, it supports managing multiple coincidences that may indeed happen in some application contexts. 
Analogously, the overall coincidence $C_{k}(j|i)$ of all the events of class $E_{k}$ in time series $TS_{j}$ with respect to the events of class $E_{k}$ in time series $TS_{i}$ is computed by taking the average coincidence of each event $y$ in time series $TS_{j}$ with respect to all events in time series $TS_{i}$ and then taking the sum of averages:

\begin{equation}
{C}_{k}(j|i) = 	
\sum\limits_{y=1}^{m_{j,k}}{
	\frac{1}{m_{i,k}} 
	\bigg[ \sum\limits_{x=1}^{m_{i,k}}{
		{c}_{k}(y,x)
	} \bigg]
}
\label{S1b}
\end{equation}

\noindent
Pairwise synchronization of the events of class $E_{k}$ for the pair of time series $\langle TS_{i}\ ;\ TS_{j}\rangle$ is computed as:

\begin{equation}
{S}_{k}(i, j) = \frac{{C}_{k}(i|j) + {C}_{k}(j|i)}{m_{i,k}+m_{j,k}} \quad S_k (i, j) \in [0,1]
\label{pairwise-intra}
\end{equation}

\noindent
Having defined the set $P \equiv \binom{T}{2}$ of all the 2-combinations of the set $T$ (i.e., each element $p \in P$ is a distinct pair $\langle TS_{i}\ ;\ TS_{j}\rangle$, with $TS_{i}$, $TS_{j} \in T$, $i \neq j$), the overall synchronization for the events of class $E_{k}$ is finally obtained as:

\begin{equation}
Q_{k} =\frac{1}{|P|}
{\sum\limits_{p \in P}{}{S_{k}(i, j)}} \quad Q_k \in [0,1]
\label{overall-intra}
\end{equation}
where the cardinality $|P|$ of set $P$ is given by the number of 2-combinations of $T$, that is:
\begin{equation}
|P| = \binom{N}{2} = \frac{N!}{2!(N-2)!}
\label{setOfcomb}
\end{equation}

\subsection{Global intra-class synchronization}
\label{subsec:global}
To compute a global synchronization index $SI$ for the events of all classes, we define the multi-class synchronization vector $\vec{Q}$ as:

\begin{equation}\label{equation:global1}
\vec{Q} = [	Q_{1},...,Q_{K} ]
\end{equation}

\noindent
$SI$ is obtained as a function of $\vec{Q}$, i.e., $SI = f(\vec{Q})$. A straightforward choice for $f$ is the average over the $K$ components of $\vec{Q}$. If event classes have e.g., different priorities, 
a set of weights $\vec{W} = [W_{1} ,...,W_{K}]$ can be associated to each class of events 
and a weighted synchronization index is computed as:

\begin{equation}\label{equation:global2}
SI_W =  f{(\vec{W},\vec{Q})}
\end{equation}

%\begin{figure}
%	\centering
%	\includegraphics[width=1\linewidth]{pictures/first_modality.pdf}
%	\caption{Computation of intra-class synchronization for each class of events.}
%	\label{fig:legssync}
%\end{figure}

\subsection{Inter-class synchronization}
\label{subsec:inter}
Given the set of event classes $E$, we may compute inter-class synchronization, i.e., the synchronization between events that
%do not belong to the same class $E_{k}$, but
belong to the different classes $E_{k1}$ and $E_{k2}$.

For each couple of time series $\langle TS_{i}\ ;\ TS_{j}\rangle$, $TS_{i}$, $TS_{j} \in T$, $i \neq j$, the coincidence between an event $x$ found in the first time series $TS_i$ and another event $y$ found in the second time series $TS_{j}$ measured by releasing the constraint that they belong to the same event class $E_{k}$, i.e., $x$ belongs to class $E_{k1}$ and $y$ to class $E_{k2}$ ($E_{k1}$ and $E_{k2}$ $\in E$). The measure of how much events $x$ and $y$ are close in time is computed within a certain interval ${\tau_{k1,k2}}$ that may depend on the considered pair of classes of events.
The temporal distance $d$ between events $x$ and $y$ (see Equation \ref{dist}) is reformulated as: 

\begin{equation}
d(x,y) = |{t^{i,k1}_{x}} - {t^{j, k2}_{y}}| 
\end{equation}
\\
Accordingly, the relative amount of coincidence $c_{k1,k2}$ (see Equation \ref{amountOfCoinc}) becomes:

\begin{equation}
{c}_{k1,k2}(x,y)  = 
\begin{cases} 
1- \frac{d(x,y)}{\tau_{k1,k2}} & \mbox{if \quad} 0 \leq d(x,y) \leq \tau_{k1,k2} \\ 
0 & \mbox{otherwise}
\end{cases}
\end{equation}

\noindent

For a pair of classes $E_{k1}$ and $E_{k2}$, the overall coincidence $C_{k1,k2}(i|j)$ of all the events of class $E_{k1}$ in time series $TS_{i}$ with respect to the events of class $E_{k2}$ in time series $TS_{j}$ and analogously, the overall coincidence $C_{k1,k2}(j|i)$ of all the events of class $E_{k1}$ in time series $TS_{j}$ with respect to the events of class $E_{k2}$ in time series $TS_{i}$ are computed by:

\begin{equation}
{C}_{k1,k2}(i|j) = 	
\sum\limits_{x=1}^{m_{i,k1}}{
	\frac{1}{m_{j,k2}} 
	\bigg[ \sum\limits_{y=1}^{m_{j,k2}}{
		{c}_{k1,k2}(x,y)
	} \bigg]
}
\label{S1-inter}
\end{equation}

\begin{equation}
{C}_{k1,k2}(j|i) = 	
\sum\limits_{y=1}^{m_{j,k2}}{
	\frac{1}{m_{i,k1}} 
	\bigg[ \sum\limits_{x=1}^{m_{i,k1}}{
		{c}_{k1,k2}(y,x)
	} \bigg]
}
\label{S2-inter}
\end{equation}

\noindent
Finally, inter-class pairwise synchronization of the events of class $E_{k1}$ and events of class $E_{k2}$, for the pair of time series $\langle TS_{i}\ ;\ TS_{j}\rangle$ is computed as:

\begin{equation}
{S}_{k1,k2}(i, j) = \frac{{C}_{k1,k2}(i|j) + {C}_{k1,k2}(j|i)}{m_{i,k1}+m_{j,k2}} 
\label{S3-inter}
\end{equation}

\noindent
and the overall synchronization for the pair of class $E_{k1}$ and $E_{k2}$ (refer to Equation \ref{pairwise-intra}):

\begin{equation}
Q_{k1,k2} =\frac{1}{|P|}
{\sum\limits_{p \in P}{}{S_{k1,k2}(i, j)}} \quad Q_{k1,k2} \in [0,1]
\label{overall-inter}
\end{equation}

\noindent
where $P$ is the set of all the 2-combinations of the set $T$ (refer to Equation \ref{setOfcomb}), and ${S}_{k1,k2}(i, j)$ and $Q_{k1,k2}$ both belong to $[0,1]$.

\section{Macro classes and Macro events} \label{sec:manipulations}
%The MECS algorithm takes as inputs a set of event classes $E$ and a set of time series $T$.
%and it is independent on how event classes are defined and how events are identified.
%In this section we will show how, by performing simple operations on the event class set, we can increase the number of modeled cases.
%\noindent
Compared with previous algorithms (\cite{quiroga2002event,kreuz2009measuring,iqbal}, see Section \ref{sec:beyond}) MECS is characterized by two important extensions: \textit{Macro classes} and \textit{Macro events}. The first one introduces the possibility to regroup the classes and compute the synchronization on different levels of abstraction corresponding to a hierarchical organization of the classes. The second extension permits to compute the synchronization between aggregations of events belonging to different classes. 

\subsection{Macro Classes}
Let's define as $Pow(E)$ the power set of the event classes set $E$ minus the empty set, i.e., $Pow(E)= \mathcal{P}(E)/\emptyset$. 
\\$Pow(E)$ has cardinality $2^K -1$. For example, if $E=\{E_{1}, E_{2}, E_{3}\} $,  $Pow(E)$  will contain the following elements:
\begin{equation}
\begin{split}
Pow(E) = &\{\{E_{1}\}, \{E_{2}\}, \{E_{3}\}, \{E_{1}, E_{2}\},  \\
& \{E_{1}, E_{3}\}, \{E_{2}, E_{3}\}, \{E_{1}, E_{2}, E_{3}\} \}
\end{split}
\end{equation}

\indent
We then define $\dot{E}\subseteq{Pow(E)}$ e.g., $\dot{E} =\{\{E_{1}, E_{2}\}, \{E_{1}, E_{2}, E_{3}\} \}$) and we consider $\dot{E}$ as the new set of event classes, i.e., in the presented example, $\dot{E_{1}} = {\dot{E_{1}},\dot{E_{2}}}$ with $\dot{E_{1}} = \{E_{1}, E_{2}\}$ and $\dot{E_{2}} = \{E_{1}, E_{2}, E_{3}\})$. 

It is possible to take $2^K -1$ subsets of the original elements of $E$, combining classes and merging them in \textit{macro classes}.
Each generated macro class is actually a single class, or the combination of two or more classes of $E$.
MECS consider each item of each set in $\dot{E}$ as belonging to the same class. Synchronization is computed using the set of equations explained in Section \ref{subsec:intra} and Section \ref{subsec:inter} by using $\dot{E}$ as input event class set. 

It is important to note that events that belong to one of the original $K$ classes can belong to more than one macro class e.g., events of class $E_{1}$ may belong to both $\dot{E_{1}}$ and $\dot{E_{2}}$ in $\dot{E}$.
%\begin{figure}
%	\centering
%	\includegraphics[width=1\linewidth]{pictures/sequences.pdf}
%	\caption{Sequences example}
%	\label{fig:sequences}
%\end{figure}
%\begin{figure*}
%	\centering
%	\includegraphics[width=1\linewidth]{pictures/manykernels.pdf}
%	\caption{Examples of a different kernel functions. From left to right: Linear, Exponential, Sigmoid an Gaussian}
%	\label{fig:manykernels}
%\end{figure*}
\begin{figure*}[h]
	\centering
	\includegraphics[height=8cm,width=1\linewidth]{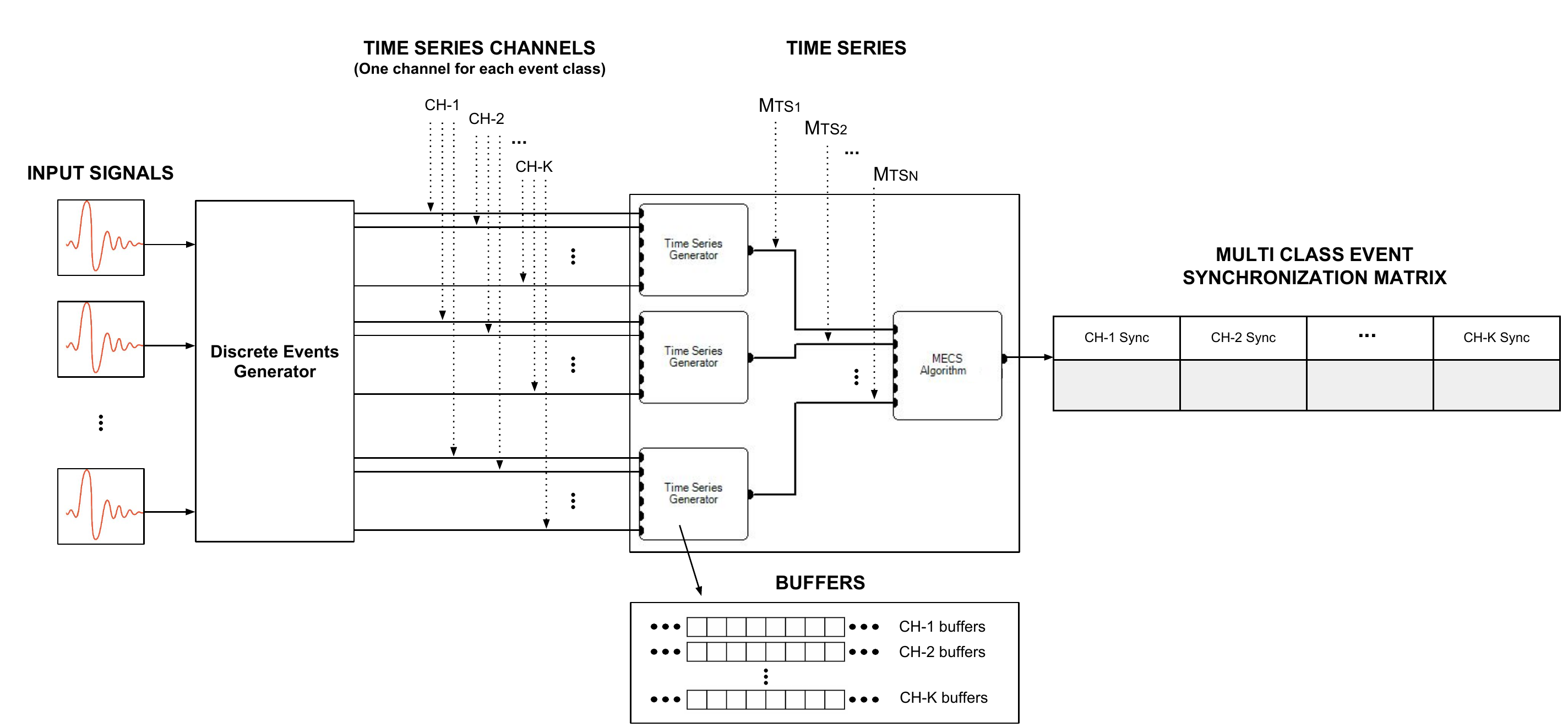}
	\caption{The MECS algorithm computing the degree of synchronization of K classes of events that have been extracted from N input signals.}
	\label{fig:esync_impl}
\end{figure*}

\subsection{Macro events}\label{sec:macro_events}
%Suppose we were able to split the motor tasks into smaller units. Each unit is therefore a sequence of movements to be followed, until completion. Given this scenario, as follow-up study, we want to determine if there is a coordination between such sequences of movements (when executed correctly) and if the completion time is synchronized across all users.
Events can be grouped into \textit{macro-events} i.e., aggregations defined by a set of constraints. An example of a macro-event is a sequence of events, where the constraint to be satisfied is the order of occurrence of each event in the sequence. 
By considering again $Pow(E)$, a sequence $S$ is defined as an ordered $n$-tuple of elements (with repetitions) where each element is referred by a sequence index: $(1,2,3,...,s-1,s)$
Starting from the elements of $E$, examples of sequences are $S_1 = \{E_{2},E_{1},E_{3}\}$, $S_2 = \{ E_{1},E_{1},E_{3},E_{1}\}$ , $S_3 = \{E_{2},E_{1}\}$ and so forth.

Let's define the following quantities:
\begin{enumerate}
	\item $S[i]$: i-th element of sequence $S$ that is an event class i.e., $S[i] \in E$,  $\forall i \in{1,...,s}$
	\item $t^{n,S[i]}_{e}$: the occurrence time of a generic event $e$ found in the time series $TS_n$ and that belongs to the $i-th$ class of sequence $S$ i.e., any $t^{n,S[i]}_{e}$ with $e \in 1,...,m_{S[i]}$.
	\item $IEI$: the \textit{Inter-Event Interval} i.e. the maximum time allowed between two events that belongs to two consecutive classes $S[i]$ and $S[i+1]$ of sequence $S$ not to interrupt the sequence.
\end{enumerate}

Then, within a time series $TS_n$, a particular sequence $S$ is detected if the following three conditions are true:

$\forall i \in{1,...,s}$
\begin{enumerate}
	%[label=(\alph*)]
	\item   $t^{n,S[i+1]}_{e} > t^{n,S[i]}_{e}$
	\item   $t^{n,S[i+1]}_{e} - t^{n,S[i]}_{e} \leq{IEI} $
	\item   no other $t^{n,l}_{e}$ occurs in [$t^{n,S[i+1]}_{e}$, $t^{n,S[i]}_{e}$] where $l\in{S}$ 
\end{enumerate}

When a sequence $S$ is detected within a time series $TS_n$, it will be treated as an event belonging to a new class named $E_{S}$.
Since sequences allows repetitions of the same elements, it is possible to define an \textit{infinite} number of sequences. Let us call $\dot{S}$ the set of all the sequences defined starting by $E$.
%with cardinality of $|\dot{(S)}|$ $\in[1,\infty]$.

The synchronization degree between sequences is computed using the set of equations explained in Section \ref{subsec:intra} with $\dot{S}$ used as input event class set.

\section{MECS Implementation}\label{sec:impl}

In this section we present an implementation of MECS algorithm. 
The schema represented in Figure \ref{fig:esync_impl} provides a graphical representation of a sample MECS application. Individual elements are shown with their interrelations. 
%Figure \ref{fig:esync_impl} shows how MECS provides an output synchronization matrix $Q$ given a set of input signals.

For simplicity, we assume that all the input signals are being generated at the same time and with a fixed frequency. 
Input signals are sampled and streamed to the \textit{Discrete Events Generator} module, that:
\begin{itemize}
	\item identifies the presence of events in the input streams and dispatches them into classes (through event detection techniques).
	\item generates $K$ discrete output streams $CH_1,...,CH_K$ called \textit{Channels} and forward them to the \textit{Time Series Generators} modules. Each channel $CH_i$ correspond to a single event class of the set $E$ (see Section \ref{sec:mecs}).
\end{itemize}

\noindent
Event detection and differentiation techniques are deliberately undefined because they strictly depends on each specific application context. 
From the The Discrete Events Generator events are sent to the Time Series Generators. 
Each \textit{Time Series Generators} module performs the following actions:
\begin{itemize}
	\item for each channel $CH_i$ fills a buffer $B_i$ with $buffDim$ samples taken from the channel streams. 
	\item fills an internal matrix $M_{TS_i}$ ($K$ rows and $buffDim$ columns) with the produced $K$ buffers, where $i \in {0,..,N}$.
	\item forwards the $TS_i$ matrix to the MECS algorithm module.
\end{itemize}

Finally, before explaining the MECS algorithm module, let us first introduce a set of auxiliary data structures: 
\begin{itemize}
	\item $CH_i$: data streams used by the \textit{Discrete Events Generator} that identifies events, and dispatches them correctly, i.e., $CH_i$ contains events of class $E_i$.
	\item $Buffer$ $B$: represents a single portion of data. During the execution of the algorithm, each channel stream is divided into buffers of size $buffDim$. 
	\item $M_{TS_i}$: matrix of channels and samples. Namely each $M_{TS_i}$ has $K$ rows and $buffDim$ columns. The value of each element of the matrix determines the presence (value $\neq 0$) or absence (value $=0$) of an event.
	\item $ECM(channel,ts)$ or Event Class Matrix: stores all the absolute positions of all the detected events. Each element of the matrix is a list of positions.
	\item $Sync(pair,channel)$ and $Tot_{Sync}(pair,channel)$ are internal data structures used to store the values of ${C}_{k}(i|j)$ and ${S}_{k}(i,j)$.
\end{itemize}
\noindent
The $ECM$, $Sync$ and $Tot_{Sync}$ data structure are re-initialized every time a new buffer arrives.
\begin{algorithm}[H]
	\KwIn{TODO mecs input}
	\KwOut{TODO mecs output}

	\For{each new $Buffer$ $B$}{ 
		Init()\;
		Compute()\;
		Finalize()\;
		$n_{buffers}$++\;
	}
	\label{alg:mecs}
	\caption{MECS computation}
\end{algorithm}

\begin{figure}
	\centering
	\includegraphics[width=1\linewidth]{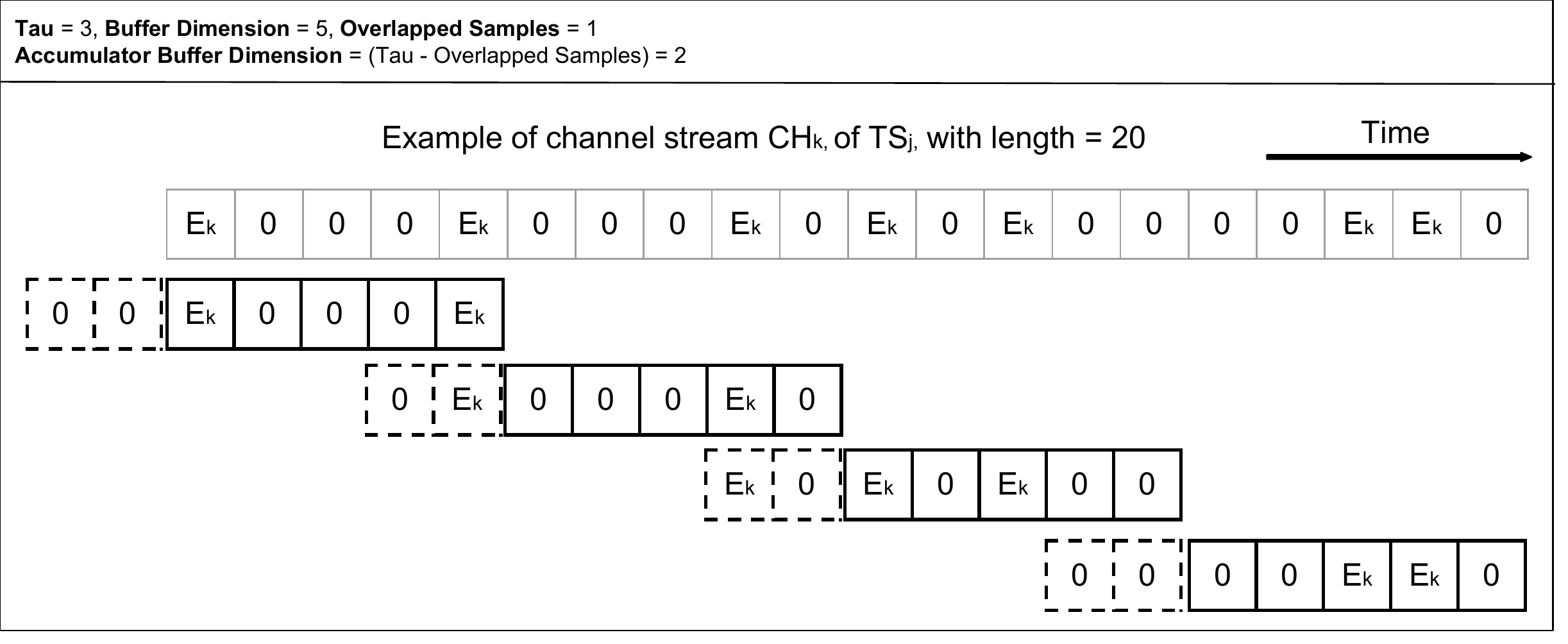}
	\caption{Use of the \textit{Accumulator buffer} structure.}
	\label{fig:merged}
\end{figure}
\begin{figure}
	\centering
	\includegraphics[width=1\linewidth]{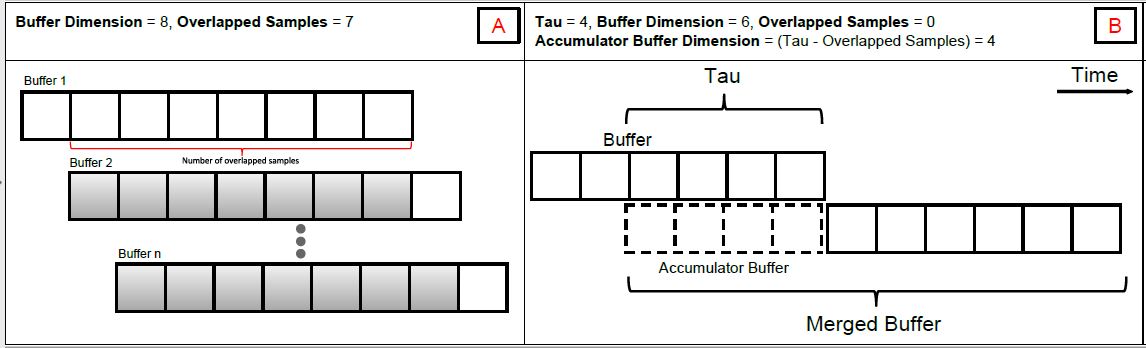}
	\caption{
		(A) Example of consecutive buffers with $buffDim$ = 8 and $n_{overlapped}$ = 7.
		(B) $Tau=4$, $n_{overlapped} =0 \implies$ \textit{Merged Buffer} is bigger than the original buffer.}
	\label{fig:buffers1}
\end{figure}
\begin{figure}
	\centering
	\includegraphics[width=1\linewidth]{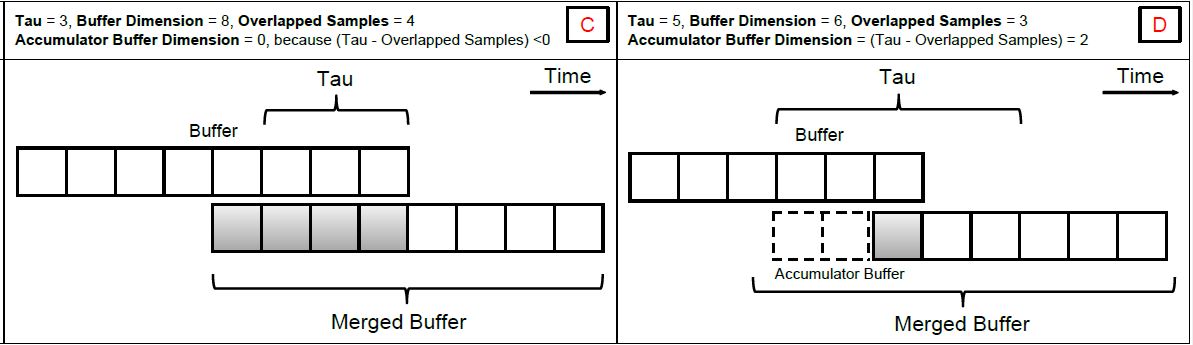}
	\caption{
		(C) $Tau< n_{overlapped} \implies$ \textit{Merged Buffer} has the same dimension of the original buffer.
		(D) $Tau > n_{overlapped} \implies$ \textit{Merged Buffer} is bigger than the original buffer.}
	\label{fig:buffers2}
\end{figure}

\subsection{Initialization}
In Algorithm \ref{alg:mecs}, the body of the main routine of MECS is reported. The main routine runs at every received buffer.
To correctly compute the synchronization vector, at each execution cycle, the algorithm stores a pre-fixed number of samples taken from the last processed buffer. Such samples are stored in a support data structure called \textit{Accumulator buffer}.
The dimension of \textit{Accumulator buffer} is computed according to $Tau$ and on the number of overlapped samples $n_{overlapped}$. Possible cases are reported in Figures \ref{fig:buffers1} and \ref{fig:buffers2}.

\newpage
\begin{algorithm}[H]
	\KwIn{TODO mecs input}
	\KwOut{TODO mecs output}
	$MergBuff = createMergedBuffer(B)$\;
	\For{$k\in\{CH_1,...,CH_K\}$}{ 
	\For{$ts\in\{M_{TS_1},...,M_{TS_N}\}$}{ 
	\For{all $sample \neq 0 \in{MergBuff}$}{
		$abs_{Pos} = n_{buffers} * off_{set} + rel_{Pos}$\;
		$ECM(k,ts).insert(abs_{Pos})$\;
	}}}
	\label{alg:init}
	\caption{MECS: INIT}
\end{algorithm}

Next, the content of the \textit{Accumulator buffer} is concatenated to the next input buffer as shown in Figure \ref{fig:merged} resulting in a \textit{Merged Buffer} of size $mergDim$.

The initialization function \emph{Init()} fills the $ECM$ (\textit{Event Class Matrix}) with all the positions of all the events found in all the available channels. Absolute positions $abs_{Pos}$ represents the occurrence timings of the events in the whole period of execution. To compute the absolute position $abs_{Pos}$ of the detected events, the MECS algorithm uses the following quantities:
\begin{itemize} 
	\item the number of received buffers $n_{buffers}$
	\item the original dimension of the buffers $buffDim$
	\item the dimension of the accumulator buffer $accDim$
	\item the number of overlapping samples $n_{overlapped}$
	\item the relative position of an event $relPos$ (the position of the sample in the buffer $B$). 
\end{itemize}

The absolute position of each event $abs_{Pos}$ is computed by using the number of received buffers ($n_{buffers}$) and an $offset$ equals to the difference between $mergDim$ and $n_{overlapped}$, as explained in Algorithm \ref{alg:init}.

\subsection{Execution}

\begin{algorithm}[H]
	\KwIn{TODO mecs input}
	\KwOut{TODO mecs output}
	$MergBuff = createMergedBuffer(B)$\;
	\For{$k\in\{CH_1,...,CH_K\}$}{ 
		\ForAll{$couples$ of time series $\langle tsi$,$tsj\rangle$ with $tsi \neq tsj$}{ 
			\ForAll{$x \in ECM[k][tsi]$}{
				\ForAll{$y \in ECM[k][tsj]$}{
				$d$ = ComputeDist$(x,y)$\;
				$sync$ = $C_{k}(i|j)$\;
				$couple$ = $\langle tsi$,$tsj\rangle$\;
				$Sync[couple,k]$ += $sync$\;
	}}}}
	\label{alg:compute}
	\caption{MECS: COMPUTE}
\end{algorithm}

The \textit{Compute()} routine calculates the distances between all the events found in all the possible pairs of time series, using the absolute positions stored in $ECM$, and saves the results in the $Sync(pair,channel)$ matrix. When this routine completes its execution, the $Sync$ matrix stores the total contribution to synchronization for each pair of time series $\langle TS_i,TS_j\rangle$ and for each event class $k$.

The \textit{Finalize()} routine performs the following steps: 
\begin{itemize} 
	\item computes pairwise synchronization of the events of class $k$ for each pair of time series $\langle TS_i,TS_j\rangle$. 
	\item computes the overall synchronization for each class $k$ dividing pairwise synchronizations ${S}_{k}$ by the set of all the 2-combinations of the considered time series.
\end{itemize} 

\begin{algorithm}[H]
	\KwIn{TODO mecs input}
	\KwOut{TODO mecs output}
	\For{$k\in\{CH_1,...,CH_K\}$}{ 
		\ForAll{$couples$ of time series $\langle tsi,tsj\rangle$ with $tsi \neq tsj$}{ 
			$couple$ = $ \langle min(tsi,tsj),max(tsi,tsj)\rangle$\;
			$Tot_{Sync}[couple,k]$ = ${S}_{k}(i, j)$\;
		}
	}
	\For{$k\in\{CH_1,...,CH_K\}$}{ 
	\ForAll{$couples$ of time series $\langle tsi,tsj\rangle$ with $tsi \neq tsj$}{ 
		$couple$ = $\langle min(tsi,tsj),max(tsi,tsj)\rangle$\;
		$Q[k]$ += $ Tot_{Sync}[couple,k]$ / Comb$(N,2)$\;
		}
	}
		
	\label{alg:finalize}
	\caption{MECS: FINALIZE}
\end{algorithm}

%\begin{algorithm}[]
%	\caption{Finalize():}
%	\begin{algorithmic}[1]	
%		\For{$k\in\{CH_1,...,CH_K\}$}
%		\ForAll{$couples$ of time series $\langle tsi,tsj\rangle$ \\ \quad \quad with $tsi \neq tsj$} 
%		\State{$couple$ = $ \langle min(tsi,tsj),max(tsi,tsj)\rangle$}
%		\State{$Tot_{Sync}[couple,k]$ = ${S}_{k}(i, j)$}
%		\EndFor
%		\EndFor		
%		\For{$k\in\{CH_1,...,CH_K\}$}
%		\ForAll{$couples$ of time series $\langle tsi,tsj\rangle$ \\ \quad \quad with $tsi \neq tsj$} 
%		\State{$couple$ = $\langle min(tsi,tsj),max(tsi,tsj)\rangle$}
%		\State{$Q[k]$ += $ Tot_{Sync}[couple,k]$ / Comb$(N,2)$}
%		\EndFor
%		\EndFor
%	\end{algorithmic}
%\end{algorithm}

\subsection{Computational complexity}

The core procedure of MECS is defined in Algorithm \ref{alg:compute}, which is executed for each input time series buffer.
The size of the input data of the procedure is: $N$ time series; $K$ channels per time series; $mergDim$ samples per channel.
Consequently, the complexity of the algorithm is:

\begin{equation}
K*(N*(N-1))/2)*mergDim^2
\end{equation}

\noindent
In most applications $N,K \ll mergeDim$.
Provided that, we conclude that the complexity of the algorithm is $O(mergeDim^5)$.

We tested the algorithm in the case study illustrated in Section \ref{sec:case}
on a Intel i7-6700k CPU @4 GhZ, 16 Gb, Windows 10 machine.
With an input consisting of $N=2$ time series with $K=4$ channels and $mergDim=10000$
samples the algorithm took on average 6.8 seconds over 10 executions.

\section{Application on synthetic data}
\label{sec:examples}s
We present two application examples of the MECS algorithm. 
We manually construct some signals and events sequences, and we provide them as input to the algorithm, reporting and commenting the corresponding output.

\subsection{Inter-class synchronization between two time series}
\begin{figure}
	\centering
	\includegraphics[width=1\linewidth]{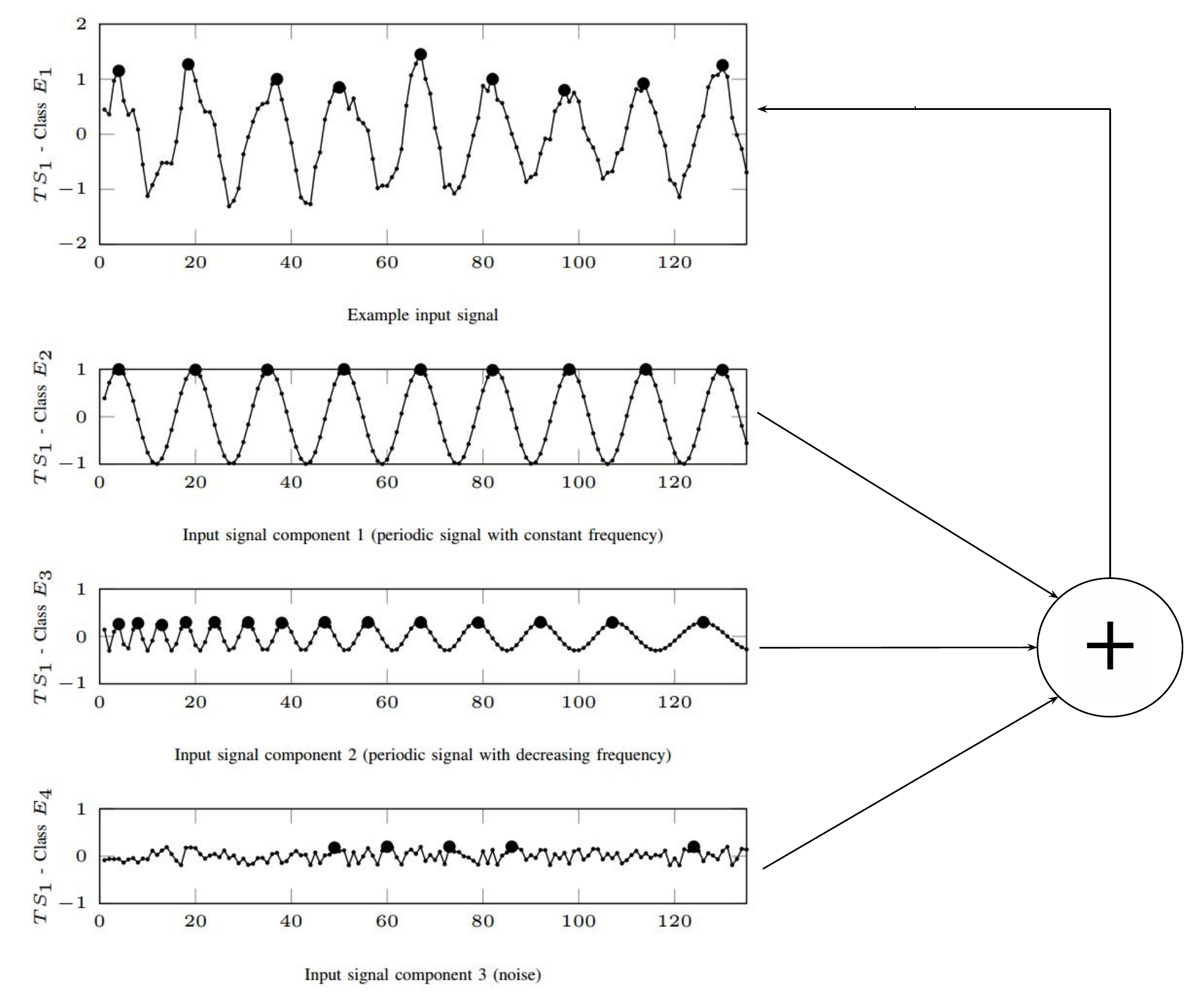}
	\caption{The input signal (highest plot) is the result of the composition of its three components: a sine wave with constant frequency and amplitude (second plot from top); a sine wave with constant amplitude and decreasing frequency (third plot from top); a noise signal (lowest plot). Dots highlight the peaks (automatically detected) of the four signals.}
	\label{fig:example-single-input}
\end{figure}

We assume that the input signal under investigation (shown in the top plot of Figure \ref{fig:example-single-input}) 
is the result of the composition of three signals:
\begin{itemize}
	\item a large sinusoidal signal with constant frequency (the second plot from top), 
	\item a smaller sinusoidal signal with decreasing frequency (the third plot from top), 
	\item  and a noise component (the fourth plot from top).
\end{itemize}

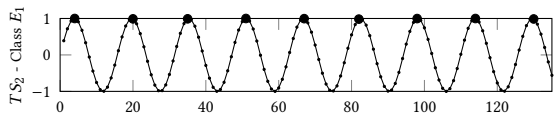
\begin{figure}[h]
	\centering
	\begin{tikzpicture}
	\begin{axis}[
	enlargelimits=false,
	width=3.2in,
	height=1in,
	ytick={-1,0,1},
	xmin=0,
	ymin=-1,
	ymax=1,
	ylabel={$TS_2$ - Class $E_1$},
	ylabel style={at={(axis description cs:0.1,0.5)},font=\tiny},
	xlabel style={font=\tiny},
	x tick label style={font=\tiny},
	y tick label style={font=\tiny}
	]
	\addplot+[
	only marks,
	color=black,mark=*,mark options={fill=black,scale=0.8}
	]
	table[x=X,y=SRP]{plots/input-SRP.csv};
	\addplot+[
	color=black,mark=*,mark options={fill=black,scale=0.2}
	]
	table[x=X,y=SR]{plots/input-single-events.csv};
	\end{axis}
	\end{tikzpicture}
	\caption{An example reference signal. The frequency and amplitude are identical to the first component of the input signal in Figure \ref{fig:example-single-input}. Dots highlight the peaks of the signal.}
	\label{fig:example-single-reference}
\end{figure}

The aim of this example is to demonstrate how the MECS algorithm can be exploited to find out the frequency of the main harmonic of a signal. 
The main harmonic is the second signal shown from the top of Figure \ref{fig:example-single-input}.
To do that, we propose to compute the synchronization of the input signal with a reference signal corresponding to the input signal main harmonic (see Figure \ref{fig:example-single-reference}).
We first extract the peaks values of the three input signal components and the reference signal: the peaks are highlighted as dots in Figure \ref{fig:example-single-input} and \ref{fig:example-single-reference}. 
Afterwards, we create 2 time-series $TS_1$ and $TS_2$:
\begin{itemize}
	\item $TS_1$ contains the events of classes $E_1$, $E_2$, $E_3$, $E_4$ corresponding to the peaks of, respectively, the input signal and three components;
	\item $TS_2$ contains the events of class $E_5$ corresponding to the reference signal peaks.
\end{itemize}

\begin{figure} [h]
	\centering
	\begin{tikzpicture}
	\begin{axis}[
	enlargelimits=false,
	width=3.5in,
	height=1in,
	ytick={0,1},
	xmin=-1,
	ymax=1,
	ymin=0,
	ylabel={$Q_{1,5}$},
	ylabel style={at={(axis description cs:0.1,0.5)},font=\tiny},
	xlabel={Reference signal Vs. Input signal},
	xlabel style={font=\tiny},
	x tick label style={font=\tiny},
	y tick label style={font=\tiny}
	]
	\addplot+[
	color=black,mark=*,mark options={fill=black,scale=0.5}
	]
	table[x=X,y=O4]{plots/output-single-gaussian-kernel.csv};
	\end{axis}
	\end{tikzpicture}
	
	\begin{tikzpicture}
	\begin{axis}[
	enlargelimits=false,
	width=3.5in,
	height=1in,
	ytick={0,1},
	xmin=-1,
	ymax=1,
	ymin=0,
	ylabel={$Q_{2,5}$},
	ylabel style={at={(axis description cs:0.1,0.5)},font=\tiny},
	xlabel={Reference signal Vs. Input signal constant frequency component},
	xlabel style={font=\tiny},
	x tick label style={font=\tiny},
	y tick label style={font=\tiny}
	]
	\addplot+[
	color=black,mark=*,mark options={fill=black,scale=0.5}
	]
	table[x=X,y=O1]{plots/output-single-gaussian-kernel.csv};
	\end{axis}
	\end{tikzpicture}
	
	\begin{tikzpicture}
	\begin{axis}[
	enlargelimits=false,
	width=3.5in,
	height=1in,
	ytick={0,1},
	xmin=-1,
	ymax=1,
	ymin=0,
	ylabel={$Q_{3,5}$},
	ylabel style={at={(axis description cs:0.1,0.5)},font=\tiny},
	xlabel={Reference signal Vs. Input signal decreasing frequency component},
	xlabel style={font=\tiny},
	x tick label style={font=\tiny},
	y tick label style={font=\tiny}
	]
	\addplot+[
	color=black,mark=*,mark options={fill=black,scale=0.5}
	]
	table[x=X,y=O2]{plots/output-single-gaussian-kernel.csv};
	\end{axis}
	\end{tikzpicture}
	
	\begin{tikzpicture}
	\begin{axis}[
	enlargelimits=false,
	width=3.5in,
	height=1in,
	ytick={0,1},
	xmin=-1,
	ymax=1,
	ymin=0,
	ylabel={$Q_{4,5}$},
	ylabel style={at={(axis description cs:0.1,0.5)},font=\tiny},
	xlabel={Reference signal Vs. Input signal noise component},
	xlabel style={font=\tiny},
	x tick label style={font=\tiny},
	y tick label style={font=\tiny}
	]
	\addplot+[
	color=black,mark=*,mark options={fill=black,scale=0.5}
	]
	table[x=X,y=O3]{plots/output-single-gaussian-kernel.csv};
	\end{axis}
	\end{tikzpicture}
	\caption{Synchronization results between the 3 signal components and the input signal reported in Figure \ref{fig:example-single-input} and the reference signal reported in Figure \ref{fig:example-single-reference}. The four synchronization values are computed at the same time by applying the MECS algorithm.}
	\label{fig:example-single-output}
\end{figure}
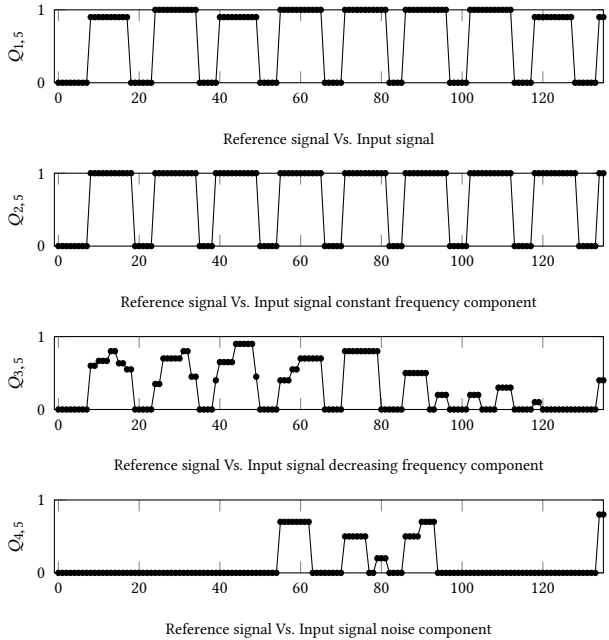

$TS_1$ and $TS_2$ are provided as input to the MECS algorithm, which computes inter-class synchronization between all the pairs of classes in $TS_1$ and $TS_2$ (that is, for each $i \in {1,2,3,4}$ it computes $Q$ between events of $E_i$ and $E_5$) and provides the values reported in Figure \ref{fig:example-single-output} as output. 
In this example, the MECS algorithm is set up with a value of $\tau=5$.
% and a linear kernel.
% (for details about kernels function see Section \ref{sec:kernel}).


As illustrated in Figure \ref{fig:example-single-output}, the synchronization between the input signal and the reference signal (the first plot from top) exhibits a rhythmical (quasi-periodic) pattern with a high constant amplitude. The same exists for the first component, which is identical to the reference signal. Conversely, the second and third components of the input signal, that is, the signal with a variable frequency and the noise component, do not exhibit the same kind of pattern.

\subsection{Macro events synchronization between two time series}
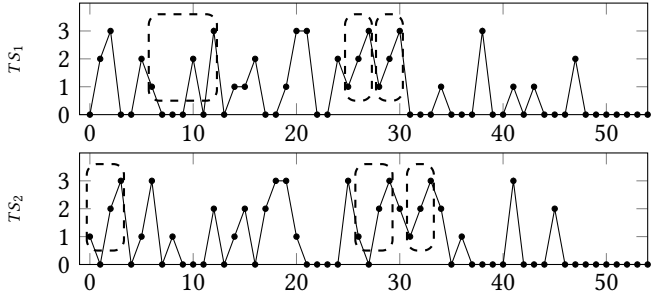
\begin{figure}[h]
	\centering
	\begin{tikzpicture}
	\begin{axis}[
	enlargelimits=false,
	width=3.6in,
	height=1.2in,
	ytick={0,1,2,3},
	xmin=-1,
	ymax=4,
	xlabel style={at={(axis description cs:0.5,0.3)},font=\tiny},
	ylabel={$TS_1$},
	ylabel style={at={(axis description cs:0.05,0.5)},font=\tiny}
	]
	\addplot+[
	color=black,mark=*,mark options={fill=black,scale=0.5}
	]
	table[x=X,y=Y1]{plots/input-sequences-linear-kernel.csv};
	\draw [black, thick, dashed, rounded corners] (axis cs:5.7,0.5) rectangle (axis cs:12.3,3.6);
	\draw [black, thick, dashed, rounded corners] (axis cs:24.7,0.5) rectangle (axis cs:27.3,3.6);
	\draw [black, thick, dashed, rounded corners] (axis cs:27.7,0.5) rectangle (axis cs:30.3,3.6);
	%	coordinates{
	%		(0,0) (1,2) (2,3) (3,0) (4,0) (5,2) (6,1) (7,0)
	%		(8,0) (9,0) (10,2) (11,0) (12,3) (13,0) (14,1) (15,1)
	%		(16,2) (17,0) (18,0) (19,1) (20,3) (21,3) (22,0) (23,0)
	%	};
	\end{axis}
	\end{tikzpicture}
	
	\begin{tikzpicture}
	\begin{axis}[
	enlargelimits=false,
	width=3.6in,
	height=1.2in,
	ytick={0,1,2,3},
	xmin=-1,
	ymax=4,
	xlabel style={at={(axis description cs:0.5,0.3)},font=\tiny},
	ylabel={$TS_2$},
	ylabel style={at={(axis description cs:0.05,0.5)},font=\tiny}
	]
	\addplot+[
	color=black,mark=*,mark options={fill=black,scale=0.5}
	]
	table[x=X,y=Y2]{plots/input-sequences-linear-kernel.csv};
	\draw [black, thick, dashed, rounded corners] (axis cs:-0.3,0.5) rectangle (axis cs:3.3,3.6);
	\draw [black, thick, dashed, rounded corners] (axis cs:25.7,0.5) rectangle (axis cs:29.3,3.6);
	\draw [black, thick, dashed, rounded corners] (axis cs:30.7,0.5) rectangle (axis cs:33.3,3.6);
	%	coordinates{
	%		(0,1) (1,0) (2,2) (3,3) (4,0) (5,1) (6,3) (7,0)
	%		(8,1) (9,0) (10,0) (11,0) (12,2) (13,0) (14,1) (15,2)
	%		(16,0) (17,2) (18,3) (19,3) (20,1) (21,0) (22,0) (23,0)
	%	};
	\end{axis}
	\end{tikzpicture}
	\caption{MECS applied to two time series. Macro events are triggered when the following sequence of events $S$ = $\{E_1,E_2,E_3\}$ is detected (see Section \ref{sec:macro_events}). The three events must occur in the specified order (in this case any amount of samples with no events can appear between them, i.e., $IEI = \infty $). Detected macro events are highlighted in the figure with dashed boxes enclosing each single event that contributes to the final sequence. 	For example, the first macro event found in $TS_1$ (in the top left corner of the figure) consists of the event of class $E_1$ followed by three samples with no events, then the event of class $E_2$, another empty sample and finally an event of class $E_3$.}
	\label{fig:example-sequences}
\end{figure}

In this second example we apply MECS to two input time series $TS_1$ and $TS_2$ consisting of events belonging to three classes (i.e., $E_1$, $E_2$, $E_3$) (see Figure \ref{fig:example-sequences}). 
Such events are represented within the time series by the corresponding positive integer numbers $1,2,3$, while, when no event is triggered, a value of zero is set. For example, $TS_2$ in Figure \ref{fig:example-sequences} starts with an event of class $E_1$, then no events are triggered for the second sample of the time series, then an event of class $E_2$ is triggered, and so on.

\begin{figure}[h]
	\centering	
	\begin{tikzpicture}
	\begin{axis}[
	enlargelimits=false,
	width=3.6in,
	height=1.3in,
	ytick={0,1},
	ymax=1,
	ymin=0,
	xlabel style={at={(axis description cs:0.5,0.3)},font=\tiny},
	ylabel={$Q_S1$},
	ylabel style={at={(axis description cs:0.05,0.5)},font=\tiny},
	]
	\addplot+[
	color=black,mark=*,mark options={fill=black,scale=0.5}
	]
	table{plots/output-sequences-linear-kernel.csv};
	\end{axis}
	\end{tikzpicture}		
	\caption{MECS algorithm output, between $TS_1$ and $TS_2$, with $\tau=20$ samples.}
	\label{figure:example2}
\end{figure}
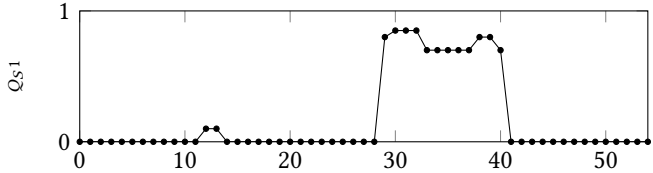

We set up the MECS algorithm to detect synchronization of macro events (sequences)  consisting of an event of class $E_1$, followed by an event of class $E_2$, followed by an event of class $E_3$ (the concept of \textit{sequence} has been introduced in Section \ref{sec:macro_events}). We chose a value of $\tau$ is $20$.
% and run the algorithm 3 times to compare different kernels: uniform, linear and Gaussian. 
The output of the algorithm, illustrated in Figure \ref{figure:example2}, is the amount of synchronization between $S$ = $\{E_1,E_2,E_3\}$ events in the two input time series. 
The Figure shows 
%that an uniform kernel (the third plot from top) provides only two possible outputs, 0 or 1, that depend on the presence or absence of event sequences within a distance of $\tau$ samples. Conversely, the linear kernel (the fourth plot from top) 
that the outputs are non-discrete values, depending on the distance.

%\begin{center}
%\end{center}
%\begin{center}
\vspace{-0.2in}
%\end{center}
\section{A Case Study}\label{sec:case}
%The research challenge that led to the development of the MECS algorithm is to quantify intra-personal and inter-personal synchronization in various human multi-modal behaviors. 
We describe an example of application of MECS to measure multi-modal, intra-personal synchronization between respiration phases and kinetic energy captured while performing body movements characterized by different expressive qualities.

Taking a breath is a physical action that can influence the body movements performed at the same time. 
Similarly, body movements expressing abrupt changes of velocity and acceleration can influence the respiration pattern. 
Rhythm of respiration synchronizes with repetitive motor activities such as running \cite{bernasconi1993,hoffmann2012}, 
or rowing \cite{bateman2006}. Moreover, respiration plays an important role in learning physical activities, such as yoga or tai-chi.
%In the proposed case study, we focus on dance to investigate the expressive qualities through the analysis of synchronization on multi-modal data.
%During a performance, dancers are used to display a large variety of expressive qualities. 

\subsection{Definitions of expressive qualities}
We analyze synchronization on multi-modal data of the dancers performing different expressive qualities. During a performance, dancers are used to display a large variety of expressive qualities. As in \cite{lussu2016}, we focus on movements displaying two particular expressive qualities: Fluidity and Impulsivity.

A \textit{fluid movement} is characterized by the following properties:  
(i) the movements of every single body joint are smooth;
%following the standard definitions  in  the  literature  of  biomechanics  [30];  
(ii) energy   is  free  to  propagate along the kinematic chains of the body (e.g., from head to trunk, from shoulders to arms) according to a
coordinated wave-like propagation \cite{alborno2016,Piana2016}.

An \textit{impulsive movement} is characterized by a sudden and non-predictable change of velocity, which is usually produced without exhibiting a preparation phase \cite{niewiadomski2015}.
Examples of impulsive movements are: avoidance movements (e.g., when hearing a sudden and unexpected noise) and movements made to recover from a loss of balance. That is, quick and repetitive
movements are not impulsive. 

While in classic dance (e.g., classic ballet) fluidity is a major expressive quality, impulsive movements are more important for contemporary dance artists (see for example the choreographies of Sagi Gross's Dance Company\footnote{\url{http://www.grossdancecompany.com}}). 

\subsection{Hypothesis}

\begin{figure}
	\centering
	\includegraphics[scale=0.3]{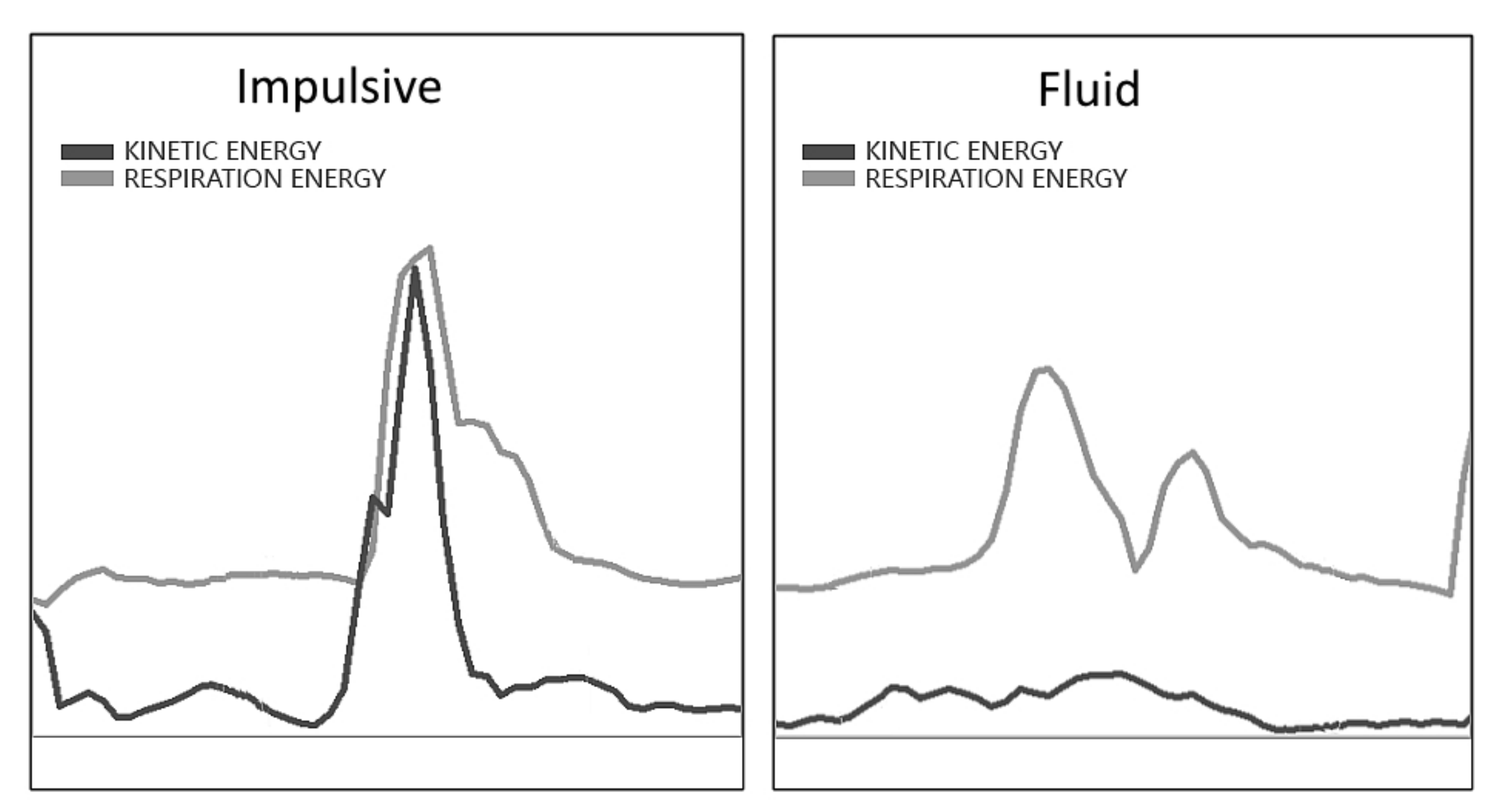}\\
	A \hspace{4cm} B
	\caption{Two parts of the same dance performances: A) impulsive movements and B) fluid movements.}
	\label{fig:eywSignal}
\end{figure}

Imagine a dancer mostly dancing fluidly and suddenly displaying one or more very impulsive movements. 
Figure \ref{fig:eywSignal} shows two examples of multi-modal data that may be captured in such a situation. Figure \ref{fig:eywSignal}-A is an example of impulsive movement, and Figure \ref{fig:eywSignal}-B is a fluid movement performed by the same dancer. 
The two signals refer to \textit{movement energy}, computed as the kinetic energy of the whole body, and to  \textit{respiration}, computed as the energy of the audio signal captured by a microphone (that was located near to the dancer's mouth).  
Energy peaks of the audio respiration signal and those of the kinetic energy are closer in Figure \ref{fig:eywSignal}-A than in Figure \ref{fig:eywSignal}-B. This is reasonable, as we may expect that different modalities may be more synchronized at the time of an impulse.

The respiration rhythm is interrupted by the impulse: a new respiration phase starts and it is synchronized with the kinetic energy peak caused by a sudden increase of velocity.
Formally, we hypothesize that during an impulsive movement there is a single strong peak in the respiration signal that appears immediately after the beginning of the respiration phase and which is synchronized with a similar peak in kinetic body energy. 
Such coincidence may depend on the respiration phase and it is not observed for fluid movements.

\subsection{Input}
We applied the MECS algorithm on 90 seconds of a recordings of a female professional dancer (see \cite{piana} for more details about the recording setup). 
The dancer was asked to move fluidly most of the time, and to perform several impulsive movements in between. The data consist of audio of the respiration captured using a wireless microphone (mono, 48kHz), two accelerometers (xOSC \footnote{\url{http://x-io.co.uk/x-osc}}) placed on the arms of the dancer, and 2 video cameras (1280x720, at 50fps).

\subsection{Features extraction}
\subsubsection{Respiration Energy (RE)}
The audio stream was segmented in frames of 1920 samples.
The instantaneous energy of the audio signal was computed on each single frame using Root Mean Square (RMS).
Next, we extracted the envelope of the instantaneous audio energy 
using an 8-frames buffer.

\subsubsection{Kinetic Energy (KE)}
Kinetic energy was computed from the data captured by the two accelerometers placed on the dancer's arms. 
Velocity was obtained by integrating the values obtained from the accelerometers. 
Next, we computed the average kinetic energy by taking the mean value of the kinetic energies obtained from the data coming from the two accelerometers. 
The accelerometer signals and the kinetic energy obtained thereof were sampled at 50fps. 
To keep synchronization with the audio energy signal, every second value was taken. 

%\begin{figure*}[h]
%	\centering
%	\includegraphics[width=1\linewidth]{pictures/EQA.pdf}
%	\caption{A simplified view of the segmentation of the dance performance. The interval of each segment has been annotated manually. Some parts of the performance do not present any of the two considered qualities. Movement and respiration features (KE and RE) has been extracted on the entire performance without differentiating between the segments. Annotations are used to define the time series and the set of events used to perform the analysis of synchronization.}
%	\label{fig:schemaeqa}
%\end{figure*}

\subsection{Manual annotation}
\subsubsection{Expressive Quality Annotation (EQA)}
An expert in expressive movement qualities annotated the beginning
of each segment of the video where impulsive or fluid movements can be perceived.
$EQA$ was used to distinguish peaks and respiration phases belonging to segments of impulsive and fluid movements.

\subsubsection{Respiration Phase Annotation (EX/IN)}
Although the inspiration and expiration phases can be automatically extracted (see as an example \cite{yahya2014}) from the audio signal, to ensure high precision of the segmentation we performed manual annotation.
The audio signal was annotated by an expert who used the Audacity
software\footnote{http://www.audacityteam.org} to assign the start and end time of each 
occurrence of each inspiration and expiration phase detected in the audio signal.

\subsection{Applying MECS} \label{sec:mecsComputation}
Our aim is to understand if the synchronization between a single respiration phase's energy and the kinetic energy of the body, grows in correspondence of impulsive behaviors given the two respiration phases.

\begin{figure*}[h]
	\centering
	\includegraphics[scale=1]{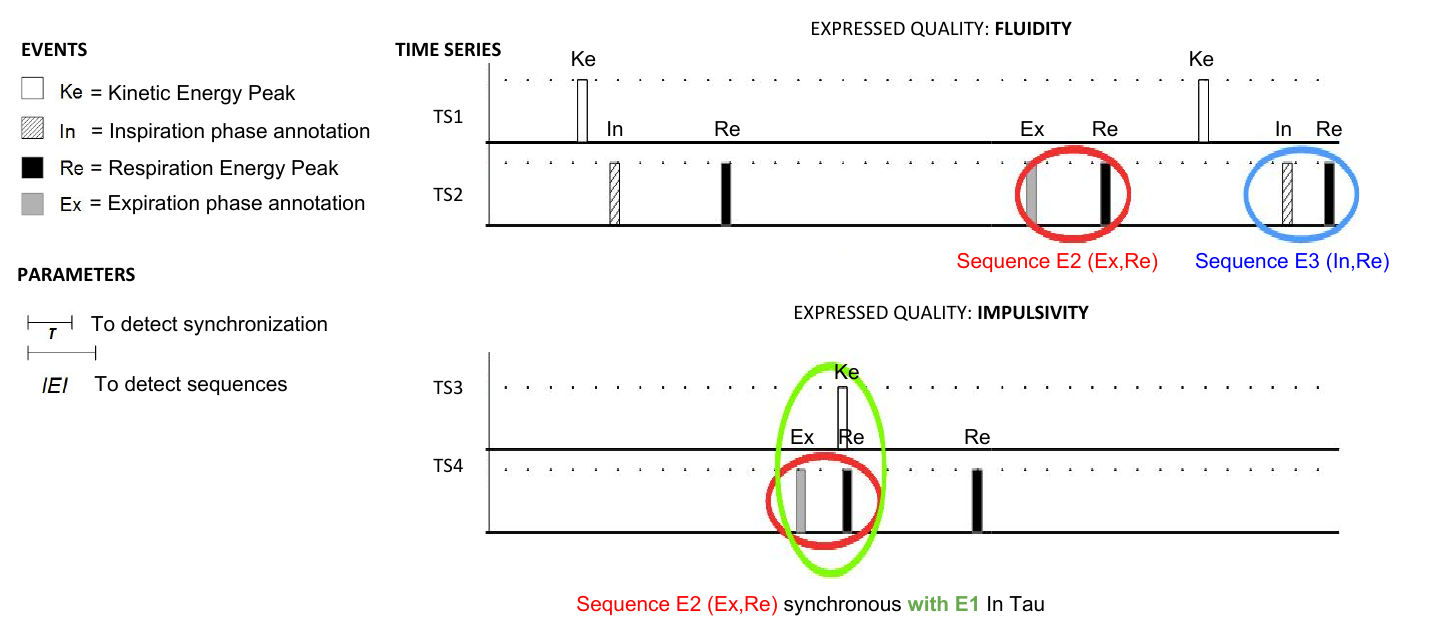}
	\caption{Synchronization between respiration and kinetic energy: the four time series of the extracted features before sequence detection.}
	\label{fig:schemasync1}
\end{figure*}

In order to apply the MECS algorithm peaks, were extracted the kinetic energy signal $KE$ and the respiration energy signal $RE$. 
Following the $EQA$ annotations, a set of $N = 4$ time series $T = \{TS_1, TS_2, TS_3, TS_4\}$ was obtained, such that:
\begin{itemize}
	\item 
	$TS_1$ contains a value different than zero in correspondence of peaks in the kinetic energy signal $KE$ during an \textbf{fluid movement}.
	\item 
	$TS_2$ contains a value different than zero in correspondence of:
	\begin{itemize}
		\item peaks in the respiration energy signal $RE$,
		\item the start time of an inspiration phase (i.e., an \textit{IN} annotation),
		\item the start time of an expiration phase (i.e., an \textit{EX} annotation) during a \textbf{fluid movement}.
	\end{itemize}
	\item 
	$TS_3$ is the equivalent of  $TS_1$ performed during an \textbf{impulsive movement}.
	\item 
	$TS_4$ is the equivalent of  $TS_2$ performed during an \textbf{impulsive movement}.
\end{itemize}	
%	\item $TS_3$ contains a value different from zero in correspondence of peaks in the kinetic energy signal $KE$ during \textbf{impulsive movement}.
%	\item $TS_4$ contains a value different from zero in correspondence of peaks in the respiration energy signal $RE$, another value different from zero in correspondence of the start time of an inspiration phase, and a further value different from zero in correspondence of the start time of an expiration phase during a \textbf{impulsive movement}.
The MECS algorithm was then applied to measure inter-class synchronization between $K = 3$ classes of events, i.e., $E = \{E_1, E_2, E_3\}$, in the time series belonging to $T$. The three classes were defined as follows:

\begin{itemize}
	\item $E_1$ is the class of the events consisting of peaks in the kinetic energy signal $KE$. These events are contained in time series $TS_1$ and $TS_3$.
	\item $E_2$ is the class of the macro-events, in the respiration energy signal $RE$, consisting of the following sequence $Seq=(IN,RE)$, i.e., the beginning of an inspiration is nearly immediately followed by a peak in the respiration energy signal $RE$. These sequences can be detected in time series $TS_2$ and $TS_4$.
	\item $E_3$ is the class of the macro-events, in the respiration energy signal $RE$, consisting of the following sequence $Seq=(EX,RE)$, i.e., the beginning of an inspiration is nearly immediately followed by a peak in the respiration energy signal $RE$. These sequences can be detected in time series $TS_2$ and $TS_4$.
\end{itemize}

To model the occurrence of a peak in the respiration energy signal $RE$ nearly immediately after 
the beginning of a respiration phase $IN$ or $EX$, we tuned the $IEI$ parameter in the conditions 
that events should satisfy for a sequence to be detected (see Section \ref{sec:macro_events}).

We ran MECS to compute: $C_1 = S_{1,2}(1,2)$, $C_2 = S_{1,3}(1,2)$,  $C_3 =  S_{1,2}(3,4)$, $C_4 = S_{1,3}(3,4)$.
%\begin{itemize}
%	\item 
%% between event classes $E_1$ and $E_2$ for 	 	
%%	fluid movements, i.e., in time series $TS_1$ and $TS_2$,
%	\item 
%% between event classes $E_1$ and $E_3$ for 	 	
%%	fluid movements, i.e., in time series $TS_1$ and $TS_2$, 
%	\item
%% between event classes $E_1$ and $E_2$ for 	 	
%%	impulsive movements, i.e., in time series $TS_3$ and $TS_4$, 
%	\item 
%% between event classes $E_1$ and $E_3$ for 	 	
%%	impulsive movements, i.e., in time series $TS_3$ and $TS_4$. 
%\end{itemize}

\noindent
%The analysis was performed separately for the four conditions $C_1$-$C_4$ on the same input data.
MECS parameters were given the following values: $\tau_{k_1,k_2} = 15$ samples (18ms) for every pair $(k_1, k_2)$ of event classes, $IEI = 25$ samples (50ms). 
%A linear kernel was used.
For each condition $C_h$, time series $TS_1$, $TS_2$, $TS_3$ and $TS_4$ were divided into $N_b$ buffers and, for each buffer, the synchronization value $S_{k_1,k_2}$ was computed, and finally it was used to compute an overall synchronization score.

%In each condition we have two classes of the events: $KE$ and corresponding sequence class $Seq()$. Thus we use the Equation \ref{S2} to compute the value $S_x^y$ of the synchronization.
%% for each respiration phase present in the trial. 

For each condition $C_1$ - $C_4$, we count the number of times inter-class pairwise synchronization $S_{k_1,k_2}(i,j) > 0$ (i.e., the number of times at least a partial synchronization is detected) and we normalize such a number by dividing it by the number of occurrences of the respiration phase events related to (i.e., inspiration of $C_1$ and $C_3$, and expiration of $C_2$ and $C_4$). That is, the final synchronization score for condition $C_h$, $h = 1...4$ is obtained as:
\noindent
\begin{equation}
\begin{aligned}
Q_{C_h} = \frac{1}{Y} \sum \limits_{t=1}^{N_b} {\Theta(S_{k_1,k_2}(i,j))} 
\end{aligned}
\end{equation}
\noindent
where: $i$ and $j$, and $k_1$ and $k_2$ are the indexes of the time series $TS_i$ and $TS_j$ and belong to the event classes $E_{k_1}$ and $E_{k_2}$ involved in condition $C_h$, respectively; $\Theta(z)$ is the Heaviside function; $Y$ is the number of occurrences of the respiration phase (inspiration or expiration) involved in condition $C_h$.

%Normalized quantity of the overall synchronization in the whole trial $TS_x^y$, $x \in \{I,F\}, y \in \{EX,IN\}$:
%
%\begin{equation}
%\begin{aligned}
%TS_x^y =  \frac {||S_x^y || : S_x^y > 0 }{ Y } 
%\end{aligned}
%\end{equation}
%
%where $|| S_x^y ||$ corresponds to the number of $y$ phases in a trial for which the at least partial synchronization occurs (i.e., the synchronization is higher than 0) and where $Y$ is the number of occurrences of phase $y$ during the movements of the quality $x$.

It is important to notice that the original ES algorithm is not powerful enough to model these complex relationship between modalities. In detail, if we want to analyze sequences of peaks in two modalities (respiration audio energy and kinetic energy), which should be treated as a single event, if and only if such peaks appear in a sufficiently ``short'' time window. 
The MECS algorithm allows us to detect sequences and to analyze their development along time. Moreover, MECS also allows us to distinguish two different classes of events related to respiration, i.e., inspiration and expiration. In this way, we are able to check whether the sequence of peaks in impulsive movements happens for any respiration phase, or it is related to one of the two respiration phases.

\subsection{Results} 
\begin{figure}
	\centering
	\includegraphics[width=1\linewidth]{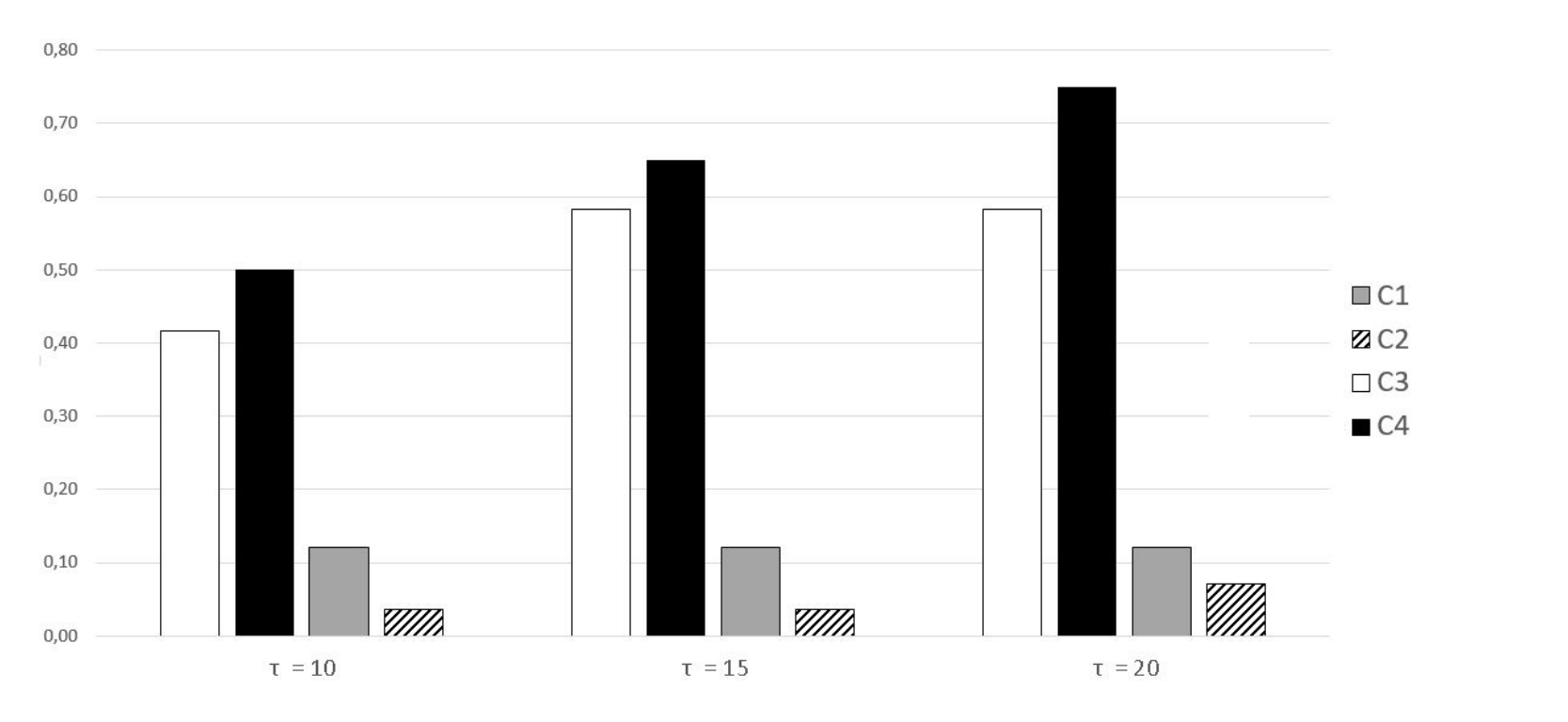}
	\caption{Synchronization results for $\tau_{k_1,k_2}$ equal to 10, 15, and 20 samples (at 50fps, the correspond to 12, 18 and 24 milliseconds). The same value is used for every pair of event classes $(k_1, k_2)$.}
	\label{fig:usecasediffernttaus}
\end{figure}

The number of inhalations and exhalations in the analyzed trial was similar (45 vs. 48). The number of inspiration intervals was lower for impulsive movements than for fluid ones (12 vs. 33). When considering the results the MECS algorithm provided for impulsive movements (conditions $C_3$ and $C_4$), the total number of the synchronized events was much higher for the expiration phases than for the inspiration phases (13 vs. 7), but the normalized score for both respiration phases is similar with $Q_{C_3} = 0.58$ and $Q_{C_4} = 0.65$. When comparing these results with those obtained for fluid segments the difference was big such that only few synchronized fluid segments were observed for the inspiration phase with $Q_{C_1} = 0.12$, and even less for the expiration phase $Q_{C_2} = 0.04$. 
Repeating the same procedure with different values of $\tau_{k_1,k_2}$ showed similar results (see Figure \ref{fig:usecasediffernttaus}). 

\section{Conclusion}\label{sec:conclusion}
In this paper, we presented Multi Event Class Synchronization (MECS), a novel algorithm for event synchronization analysis.
With respect to other techniques existing in the literature, MECS introduces the possibility to measure synchronization between different classes of events that are detected in multiple time series, providing different synchronization measures (i.e., a synchronization measure for each class, a global synchronization measure, synchronization measures obtained by combining classes).
Furthermore we described how to operate on the event set, introducing the concept of \textit{macro classes} and \textit{macro events}
%, we explained how to tune the computation of synchronization by changing the kernel function of the algorithm, 
and we  proposed a possible implementation.
Finally, we illustrated how MECS has been applied to artificially generated and real-life data.

%More applications of the algorithm would be needed to effectively highlight the real potentialities of our approach. 
%For example, MECS can be used to analyze the degree of synchronization of movements performed by multiple users (inter-personal synchronization), or to measure coordination between an user and a robotic agent.

\begin{acks}
This research has received funding from the European Union Horizon 2020 research and innovation programme.
\end{acks}

% Bibliography
\bibliographystyle{ACM-Reference-Format}
\bibliography{sample-bibliography}


\begin{thebibliography}{16}

%%% the Web address.

\ifx \showCODEN    \undefined \def \showCODEN     #1{\unskip}     \fi
\ifx \showDOI      \undefined \def \showDOI       #1{#1}\fi
\ifx \showISBNx    \undefined \def \showISBNx     #1{\unskip}     \fi
\ifx \showISBNxiii \undefined \def \showISBNxiii  #1{\unskip}     \fi
\ifx \showISSN     \undefined \def \showISSN      #1{\unskip}     \fi
\ifx \showLCCN     \undefined \def \showLCCN      #1{\unskip}     \fi
\ifx \shownote     \undefined \def \shownote      #1{#1}          \fi
\ifx \showarticletitle \undefined \def \showarticletitle #1{#1}   \fi
\ifx \showURL      \undefined \def \showURL       {\relax}        \fi
% invisible to TeX
\providecommand\bibfield[2]{#2}
\providecommand\bibinfo[2]{#2}
\providecommand\natexlab[1]{#1}
\providecommand\showeprint[2][]{arXiv:#2}

\bibitem[\protect\citeauthoryear{Alborno, Piana, Mancini, Niewiadomski, Volpe,
  and Camurri}{Alborno et~al\mbox{.}}{2016}]%
        {alborno2016}
\bibfield{author}{\bibinfo{person}{Paolo Alborno}, \bibinfo{person}{Stefano
  Piana}, \bibinfo{person}{Maurizio Mancini}, \bibinfo{person}{Radoslaw
  Niewiadomski}, \bibinfo{person}{Gualtiero Volpe}, {and}
  \bibinfo{person}{Antonio Camurri}.} \bibinfo{year}{2016}\natexlab{}.
\newblock \showarticletitle{Analysis of Intrapersonal Synchronization in
  Full-Body Movements Displaying Different Expressive Qualities}. In
  \bibinfo{booktitle}{\emph{Proceedings of the International Working Conference
  on Advanced Visual Interfaces}} \emph{(\bibinfo{series}{AVI '16})}.
  \bibinfo{publisher}{ACM}, \bibinfo{address}{New York, NY, USA},
  \bibinfo{pages}{136--143}.
\newblock
\showISBNx{978-1-4503-4131-8}
\urldef\tempurl%
\url{https://doi.org/10.1145/2909132.2909262}
\showDOI{\tempurl}


\bibitem[\protect\citeauthoryear{Bateman, McGregor, Bull, Cashman, and
  Schroter}{Bateman et~al\mbox{.}}{2006}]%
        {bateman2006}
\bibfield{author}{\bibinfo{person}{AH Bateman}, \bibinfo{person}{AH McGregor},
  \bibinfo{person}{AMJ Bull}, \bibinfo{person}{PMM Cashman}, {and}
  \bibinfo{person}{RC Schroter}.} \bibinfo{year}{2006}\natexlab{}.
\newblock \showarticletitle{Assessment of the timing of respiration during
  rowing and its relationship to spinal kinematics}.
\newblock \bibinfo{journal}{\emph{Biology of Sport}}  \bibinfo{volume}{23}
  (\bibinfo{year}{2006}), \bibinfo{pages}{353--365}.
\newblock


\bibitem[\protect\citeauthoryear{Bernasconi and Kohl}{Bernasconi and
  Kohl}{1993}]%
        {bernasconi1993}
\bibfield{author}{\bibinfo{person}{Paolo Bernasconi} {and}
  \bibinfo{person}{Jana Kohl}.} \bibinfo{year}{1993}\natexlab{}.
\newblock \showarticletitle{Analysis of co-ordination between breathing and
  exercise rhythms in man}.
\newblock \bibinfo{journal}{\emph{J. Physiol}}  \bibinfo{volume}{471}
  (\bibinfo{year}{1993}), \bibinfo{pages}{693--706}.
\newblock


\bibitem[\protect\citeauthoryear{Hoffmann, Torregrosa, and Bardy}{Hoffmann
  et~al\mbox{.}}{2012}]%
        {hoffmann2012}
\bibfield{author}{\bibinfo{person}{Charles~P. Hoffmann},
  \bibinfo{person}{G?rald Torregrosa}, {and} \bibinfo{person}{Beno?t~G.
  Bardy}.} \bibinfo{year}{2012}\natexlab{}.
\newblock \showarticletitle{Sound Stabilizes Locomotor-Respiratory Coupling and
  Reduces Energy Cost}.
\newblock \bibinfo{journal}{\emph{PLoS ONE}} \bibinfo{volume}{7},
  \bibinfo{number}{9} (\bibinfo{date}{09} \bibinfo{year}{2012}).
\newblock


\bibitem[\protect\citeauthoryear{Hung and Gatica-Perez}{Hung and
  Gatica-Perez}{2010}]%
        {hung2010}
\bibfield{author}{\bibinfo{person}{Hayley Hung} {and} \bibinfo{person}{Daniel
  Gatica-Perez}.} \bibinfo{year}{2010}\natexlab{}.
\newblock \showarticletitle{Estimating Cohesion in Small Groups Using
  Audio-Visual Nonverbal Behavior}.
\newblock \bibinfo{journal}{\emph{IEEE Transactions on Multimedia}}
  \bibinfo{volume}{12}, \bibinfo{number}{6} (\bibinfo{date}{Oct}
  \bibinfo{year}{2010}), \bibinfo{pages}{563--575}.
\newblock
\showISSN{1520-9210}
\urldef\tempurl%
\url{https://doi.org/10.1109/TMM.2010.2055233}
\showDOI{\tempurl}


\bibitem[\protect\citeauthoryear{Iqbal and Riek}{Iqbal and Riek}{2016}]%
        {iqbal}
\bibfield{author}{\bibinfo{person}{Tariq Iqbal} {and}
  \bibinfo{person}{Laurel~D. Riek}.} \bibinfo{year}{2016}\natexlab{}.
\newblock \showarticletitle{A Method for Automatic Detection of Psychomotor
  Entrainment}.
\newblock \bibinfo{journal}{\emph{IEEE Transactions on Affective Computing}}
  \bibinfo{volume}{7}, \bibinfo{number}{1} (\bibinfo{date}{Jan}
  \bibinfo{year}{2016}), \bibinfo{pages}{3--16}.
\newblock
\showISSN{1949-3045}
\urldef\tempurl%
\url{https://doi.org/10.1109/TAFFC.2015.2445335}
\showDOI{\tempurl}


\bibitem[\protect\citeauthoryear{Keltner}{Keltner}{1995}]%
        {Keltner1995}
\bibfield{author}{\bibinfo{person}{Dacher Keltner}.}
  \bibinfo{year}{1995}\natexlab{}.
\newblock \showarticletitle{Signs of appeasement: Evidence for the distinct
  displays of embarrassment, amusement, and shame}.
\newblock \bibinfo{journal}{\emph{Journal of Personality and Social
  Psychology}}  \bibinfo{volume}{68} (\bibinfo{year}{1995}),
  \bibinfo{pages}{441--454}.
\newblock


\bibitem[\protect\citeauthoryear{Kreuz}{Kreuz}{2011}]%
        {kreuz2011measures}
\bibfield{author}{\bibinfo{person}{Thomas Kreuz}.}
  \bibinfo{year}{2011}\natexlab{}.
\newblock \showarticletitle{Measures of spike train synchrony}.
\newblock \bibinfo{journal}{\emph{Scholarpedia}} \bibinfo{volume}{6},
  \bibinfo{number}{10} (\bibinfo{year}{2011}), \bibinfo{pages}{11934}.
\newblock


\bibitem[\protect\citeauthoryear{Kreuz, Chicharro, Andrzejak, Haas, and
  Abarbanel}{Kreuz et~al\mbox{.}}{2009}]%
        {kreuz2009measuring}
\bibfield{author}{\bibinfo{person}{Thomas Kreuz}, \bibinfo{person}{Daniel
  Chicharro}, \bibinfo{person}{Ralph~G Andrzejak}, \bibinfo{person}{Julie~S
  Haas}, {and} \bibinfo{person}{Henry~DI Abarbanel}.}
  \bibinfo{year}{2009}\natexlab{}.
\newblock \showarticletitle{Measuring multiple spike train synchrony}.
\newblock \bibinfo{journal}{\emph{Journal of neuroscience methods}}
  \bibinfo{volume}{183}, \bibinfo{number}{2} (\bibinfo{year}{2009}),
  \bibinfo{pages}{287--299}.
\newblock


\bibitem[\protect\citeauthoryear{Lakens and Stel}{Lakens and Stel}{2011}]%
        {lakens2011}
\bibfield{author}{\bibinfo{person}{Daniel Lakens} {and}
  \bibinfo{person}{Marielle Stel}.} \bibinfo{year}{2011}\natexlab{}.
\newblock \showarticletitle{If They Move in Sync, They Must Feel in Sync:
  Movement Synchrony Leads to Attributions of Rapport and Entitativity}.
\newblock \bibinfo{journal}{\emph{Social Cognition}}  \bibinfo{volume}{29}
  (\bibinfo{year}{2011}), \bibinfo{pages}{1--14}.
\newblock


\bibitem[\protect\citeauthoryear{Lussu, Niewiadomski, Volpe, and Camurri}{Lussu
  et~al\mbox{.}}{2016}]%
        {lussu2016}
\bibfield{author}{\bibinfo{person}{Vincenzo Lussu}, \bibinfo{person}{Radoslaw
  Niewiadomski}, \bibinfo{person}{Gualtiero Volpe}, {and}
  \bibinfo{person}{Antonio Camurri}.} \bibinfo{year}{2016}\natexlab{}.
\newblock \showarticletitle{Using the Audio Respiration Signal for Multimodal
  Discrimination of Expressive Movement Qualities}.
\newblock In \bibinfo{booktitle}{\emph{Human Behavior Understanding: 7th
  International Workshop, HBU 2016, Amsterdam, The Netherlands, October 16,
  2016, Proceedings}}, \bibfield{editor}{\bibinfo{person}{Mohamed Chetouani},
  \bibinfo{person}{Jeffrey Cohn}, {and} \bibinfo{person}{Albert~Ali Salah}}
  (Eds.). \bibinfo{publisher}{Springer International Publishing},
  \bibinfo{pages}{102--115}.
\newblock


\bibitem[\protect\citeauthoryear{Niewiadomski, Hyniewska, and
  Pelachaud}{Niewiadomski et~al\mbox{.}}{2011}]%
        {niewiadomski2011}
\bibfield{author}{\bibinfo{person}{Radoslaw Niewiadomski},
  \bibinfo{person}{Sylwia~Julia Hyniewska}, {and} \bibinfo{person}{Catherine
  Pelachaud}.} \bibinfo{year}{2011}\natexlab{}.
\newblock \showarticletitle{Constraint-Based Model for Synthesis of Multimodal
  Sequential Expressions of Emotions}.
\newblock \bibinfo{journal}{\emph{Affective Computing, IEEE Transactions on}}
  \bibinfo{volume}{2}, \bibinfo{number}{3} (\bibinfo{year}{2011}),
  \bibinfo{pages}{134--146}.
\newblock
\showISSN{1949-3045}
\urldef\tempurl%
\url{https://doi.org/10.1109/T-AFFC.2011.5}
\showDOI{\tempurl}


\bibitem[\protect\citeauthoryear{Niewiadomski, Mancini, Volpe, and
  Camurri}{Niewiadomski et~al\mbox{.}}{2015}]%
        {niewiadomski2015}
\bibfield{author}{\bibinfo{person}{Radoslaw Niewiadomski},
  \bibinfo{person}{Maurizio Mancini}, \bibinfo{person}{Gualtiero Volpe}, {and}
  \bibinfo{person}{Antonio Camurri}.} \bibinfo{year}{2015}\natexlab{}.
\newblock \showarticletitle{Automated Detection of Impulsive Movements in HCI}.
  In \bibinfo{booktitle}{\emph{Proceedings of the 11th Biannual Conference on
  Italian SIGCHI Chapter}} \emph{(\bibinfo{series}{CHItaly 2015})}.
  \bibinfo{publisher}{ACM}, \bibinfo{address}{New York, NY, USA},
  \bibinfo{pages}{166--169}.
\newblock
\showISBNx{978-1-4503-3684-0}


\bibitem[\protect\citeauthoryear{Piana, Coletta, Ghisio, Niewiadomski, Mancini,
  Sagoleo, Volpe, and Camurri}{Piana et~al\mbox{.}}{2016}]%
        {Piana2016}
\bibfield{author}{\bibinfo{person}{Stefano Piana}, \bibinfo{person}{Paolo
  Coletta}, \bibinfo{person}{Simone Ghisio}, \bibinfo{person}{Radoslaw
  Niewiadomski}, \bibinfo{person}{Maurizio Mancini}, \bibinfo{person}{Roberto
  Sagoleo}, \bibinfo{person}{Gualtiero Volpe}, {and} \bibinfo{person}{Antonio
  Camurri}.} \bibinfo{year}{2016}\natexlab{}.
\newblock \showarticletitle{Towards a Multimodal Repository of Expressive
  Movement Qualities in Dance}. In \bibinfo{booktitle}{\emph{Proceedings of the
  3rd International Symposium on Movement and Computing}}
  \emph{(\bibinfo{series}{MOCO '16})}. \bibinfo{publisher}{ACM},
  \bibinfo{address}{New York, NY, USA}, Article \bibinfo{articleno}{10},
  \bibinfo{numpages}{8}~pages.
\newblock
\showISBNx{978-1-4503-4307-7}
\urldef\tempurl%
\url{https://doi.org/10.1145/2948910.2948931}
\showDOI{\tempurl}


\bibitem[\protect\citeauthoryear{Quiroga, Kreuz, and Grassberger}{Quiroga
  et~al\mbox{.}}{2002}]%
        {quiroga2002event}
\bibfield{author}{\bibinfo{person}{Rodrigo~Quian Quiroga},
  \bibinfo{person}{Thomas Kreuz}, {and} \bibinfo{person}{Peter Grassberger}.}
  \bibinfo{year}{2002}\natexlab{}.
\newblock \showarticletitle{Event synchronization: a simple and fast method to
  measure synchronicity and time delay patterns}.
\newblock \bibinfo{journal}{\emph{Physical review E}} \bibinfo{volume}{66},
  \bibinfo{number}{4} (\bibinfo{year}{2002}), \bibinfo{pages}{041904}.
\newblock


\bibitem[\protect\citeauthoryear{Yahya and Faezipour}{Yahya and
  Faezipour}{2014}]%
        {yahya2014}
\bibfield{author}{\bibinfo{person}{Omar Yahya} {and} \bibinfo{person}{Miad
  Faezipour}.} \bibinfo{year}{2014}\natexlab{}.
\newblock \showarticletitle{Automatic detection and classification of acoustic
  breathing cycles}. In \bibinfo{booktitle}{\emph{Proceedings of the 2014 Zone
  1 Conference of the American Society for Engineering Education}}.
  \bibinfo{pages}{1--5}.
\newblock
\urldef\tempurl%
\url{https://doi.org/10.1109/ASEEZone1.2014.6820648}
\showDOI{\tempurl}


\end{thebibliography}

\end{document}